\documentclass[
 reprint,
 amsmath,amssymb,
 aps,
 prb
]{revtex4-2}

\usepackage[dvipsnames]{xcolor}

\usepackage{hyperref}
\hypersetup{
    colorlinks=true,
    linkcolor=black,
    citecolor=CadetBlue,
    filecolor=CadetBlue,      
    urlcolor=CadetBlue,
}

\usepackage{pifont}

\usepackage{upgreek}

\pdfoutput=1 % for arXiv
\usepackage{graphicx}  % needed for figures
\usepackage{dcolumn}   % needed for some tables
\usepackage{bm}        % for math
\usepackage{amssymb}   % for math
\usepackage{comment}
\usepackage{amsmath,amssymb,amsthm,dsfont,amsfonts,xfrac,braket,mathtools,bm,bbm,mathrsfs}

\usepackage{xcolor}

\newcommand{\sign}{\text{sign}}

\usepackage{comment}

\usepackage[english]{babel}
\usepackage{amsmath,amsthm,amssymb}
\usepackage{amsfonts}
\usepackage{amsopn}
\usepackage{float}

\usepackage{url}

\usepackage{graphicx}
\usepackage{dcolumn}
\usepackage{bm}
\usepackage{color}

\usepackage{float}

\newcommand{\quotationmarks}[1]{``#1''}

\begin{document}

\preprint{APS/123-QED}

\title{\textbf{Long-range spin glass in a field at zero temperature} 
}% 

\author{Maria Chiara Angelini$^{3,4}$} \author{Saverio Palazzi$^{3}$}\email{Contact author: saverio.palazzi.1997@gmail.com} \author{Giorgio Parisi$^{1,3,4,5}$}\author{Tommaso Rizzo$^{2,3,4}$}

\affiliation{$^1$International Research Center of Complexity Sciences, Hangzhou International Innovation Institute, Beihang University, Hangzhou 311115, China}
\affiliation{$^2$ISC-CNR, UOS Rome, Sapienza Universit\`a di Roma, Piazzale A. Moro 2, I-00185, Rome, Italy}
\affiliation{$^3$Dipartimento di Fisica, Sapienza Universit\`a di Roma, Piazzale A. Moro 2, I-00185, Rome, Italy}
\affiliation{$^4$INFN-Sezione di Roma 1, Piazzale A. Moro 2, 00185, Rome, Italy}
\affiliation{$^5$Nanotec-CNR, UOS Rome, Sapienza Universit\`a di Roma, Piazzale A. Moro 2, I-00185, Rome, Italy}

\date{\today}% It is always \today, today,
             %  but any date may be explicitly specified

\begin{abstract}
We compute the critical exponents of the zero-temperature spin glass transition in a field on a one-dimensional long-range model, a proxy for higher-dimensional systems. Our approach is based on a novel loop expansion within the Bethe $M$-layer formalism, whose adaptation to this specific case is detailed here. The resulting estimates provide crucial benchmarks for numerical simulations that can access larger system sizes in one dimension, thus offering a key test of the theory of spin glasses in a field.
\end{abstract}

%\keywords{Suggested keywords}%Use showkeys class option if keyword
                              %display desired
\maketitle

%\tableofcontents

\section{Introduction}

The fate of spin glass transitions in finite dimensions is still an open and debated problem since the introduction of the Edwards-Anderson model \cite{Edwards_1975}, which celebrates its 50th anniversary in 2025. The presence of an external field further complicates the problem: both the existence of the transition and the value of the lower critical dimension are topics of active discussion within the scientific community.

Several studies have employed theoretical approaches, with the Renormalization Group (RG) \cite{Parisi1988,Amit_2006,zinn2021quantum,Bellac_1991} being the standard method of investigation, both within the real-space RG \cite{gardner1984spin,parisi2001renormalization,drossel2000spin,angelini2013ensemble,angelini2015spin,monthus2015fractal,wang2018fractal} and within the momentum-space RG \cite{bray1980renormalisation,temesvari2002generic,pimentel2002spin,moore2011disappearance,parisi2012replica,temesvari2017physical,charbonneau2017nontrivial,charbonneau2019morphology,holler2020one}. Unfortunately, due to the approximated (or perturbative) nature of these results, no definitive answer can be given to the problem. On the other hand, numerical simulations \cite{baity2014three,baity2014dynamical,banos2012thermodynamic, vedula2023study, aguilar2024evidence, vedula2024evidence} provide another avenue for investigation, but finite-size effects and long equilibration times often hinder their interpretation.

Simplified versions of the model have been studied, such as the Sherrington-Kirkpatrick (SK) model \cite{sherrington1975solvable}, essentially a spin glass defined on a fully connected (FC) graph, whose solution was found using the replica method \cite{parisi1980sequence,parisi1980order} and later proven to be exact \cite{talagrand2000replica,Guerra_2002,panchenko2013sherrington}. In the temperature-field plane, the model exhibits a transition from the paramagnetic to the spin glass, full-Replica Symmetry Breaking phase, characterized by the celebrated ultrametric structure of the states. The critical line is called \quotationmarks{de Almeida-Thouless} (dAT) line \cite{deAlmeida_1978}. The peculiarity of this model is that it does not predict such a transition at zero temperature, as the system remains in the spin glass phase for any value of the external field. This feature is encoded in the divergence of the dAT line as temperature approaches zero.

A different behavior emerges on the Bethe Lattice (BL) \cite{mezard2001bethe,mezard2003cavity,Parisi_2014}, denoted in this work as a random regular graph with fixed connectivity and loop lengths that diverge logarithmically with system size. Spin glass models in a field can be solved on the BL in the paramagnetic phase by means of the cavity method \cite{Mezard_86}, which corresponds, to some extent, to the Bethe-Peierls approximation \cite{bethe1935statistical,peierls1936ising}. The result is the prediction of a transition to a low temperature (or field) spin glass phase, which is described within the one-step Replica Symmetry Breaking approximation \cite{mezard2003cavity}. Unlike the SK model, the corresponding dAT line on the BL does not diverge at zero temperature \cite{Parisi_2014}. Notably, the BL solution differs significantly from the FC one, as finite connectivity induces local fluctuations of the order parameter, making it more similar to finite-dimensional systems.

An alternative approach to studying finite-dimensional transitions, based on a topological loop perturbative expansion, has been proposed in Ref. \cite{Altieri_2017}. This method, called \quotationmarks{Bethe $M$-layer construction}, builds on the Mean-Field (MF) Bethe-Peierls solution. The idea is to construct the \quotationmarks{$M$-layer lattice} by initially creating $M$ independent copies of the finite-dimensional lattice and then randomly rewiring the edges. In any dimension, as $M \to \infty$, all observables of the model approach the BL results, leading to a phase transition governed by the MF critical exponents. At finite $M$, corrections of order $\mathcal{O}(1/M)$ arise due to the presence of topological loops in the graph. These corrections are negligible above the upper critical dimension $D_\text{uc}$ but are expected to modify the critical behavior for $D < D_\text{uc}$. Performing a $1/M$ expansion yields a diagrammatic expansion in the number of topological loops, which can be analyzed using standard RG methods to determine the upper critical dimension, $D_\text{uc}$. Applying this approach to models that share the same transition type on both the BL and FC lattices reproduces results from the standard field-theoretical loop expansion, as tested for the Ising model \cite{Angelini_2024Ising}, percolation \cite{angelini2025bethe}, and spin glass in a field at high connectivity for $T>0$ \cite{angelini2018one}. However, when the BL solution differs from the FC one or when no FC solution exists, the $M$-layer construction provides new insights. This is the case for the Random Field Ising Model at zero temperature \cite{angelini2020loop}, bootstrap percolation \cite{rizzo2019fate}, glass crossover \cite{rizzo2020solvable}, and Anderson localization \cite{baroni2024corrections}. 

The $M$-layer expansion has been applied to spin glasses in a field directly at $T=0$, where the SK model remains in the spin glass phase for any field value. In contrast, the BL exhibits a critical field separating the paramagnetic and spin-glass phases. Remarkably, this expansion identified an upper critical dimension $D_\text{uc} = 8$ \cite{angelini2022unexpected}, differing from the field-theoretical result of $D_{\text{FC,uc}} = 6$ at $T \neq 0$ \cite{bray1980renormalisation} (for which the corresponding MF model is the SK model, defined on a FC graph).

A recent paper \cite{angelini2025criticalexponentsspinglass} employs the $M$-layer construction to compute the first order $\epsilon$-expansion of the critical exponents associated with the zero temperature transition of the Edwards-Anderson model in a field in finite dimensions. Remarkably, a non-trivial fixed point is found for $D_\text{uc}<8$. In the same work, the $\epsilon$-expansion for the critical exponents of the associated one-dimensional long-range (LR) model is reported, without providing details of the computation, and it is argued that the method can be applied to the LR version. In this paper, we give all the details for the computation of the above-mentioned $\epsilon$-expansion for LR critical exponents. These results provide a valuable theoretical framework for interpreting numerical simulations, in which LR one-dimensional models serve as a practical testing ground. Having reliable analytical estimates of the exponents is crucial for facilitating data analysis and improving the accuracy of numerical extrapolations. 

The paper is organized as follows. In Sec. \ref{sec:Definition of model and observables} we define the model of interest and we describe its phenomenology; in Sec. \ref{sec:Main results} we collect the main results of the work; in Sec. \ref{sec:The method: M-layer construction} we delve into the details of the $M$-layer construction and we leave Sec. \ref{sec:Conclusions} for conclusions and perspectives.

\section{Definition of model and observables}\label{sec:Definition of model and observables}

In the following, we define the model on a one-dimensional lattice with LR interactions and describe its behavior at zero temperature. The treatment is the same as in Ref. \cite{angelini2025criticalexponentsspinglass}, but it is reported here for completeness, with the observables' definitions adapted to the case studied in this paper. Specifically, we will adapt the scaling laws, which are crucial for computing the critical exponents of the LR model.

\subsection{The model}\label{sec: the model}
\begin{figure}[H]
        \centering        \includegraphics[scale=0.4]{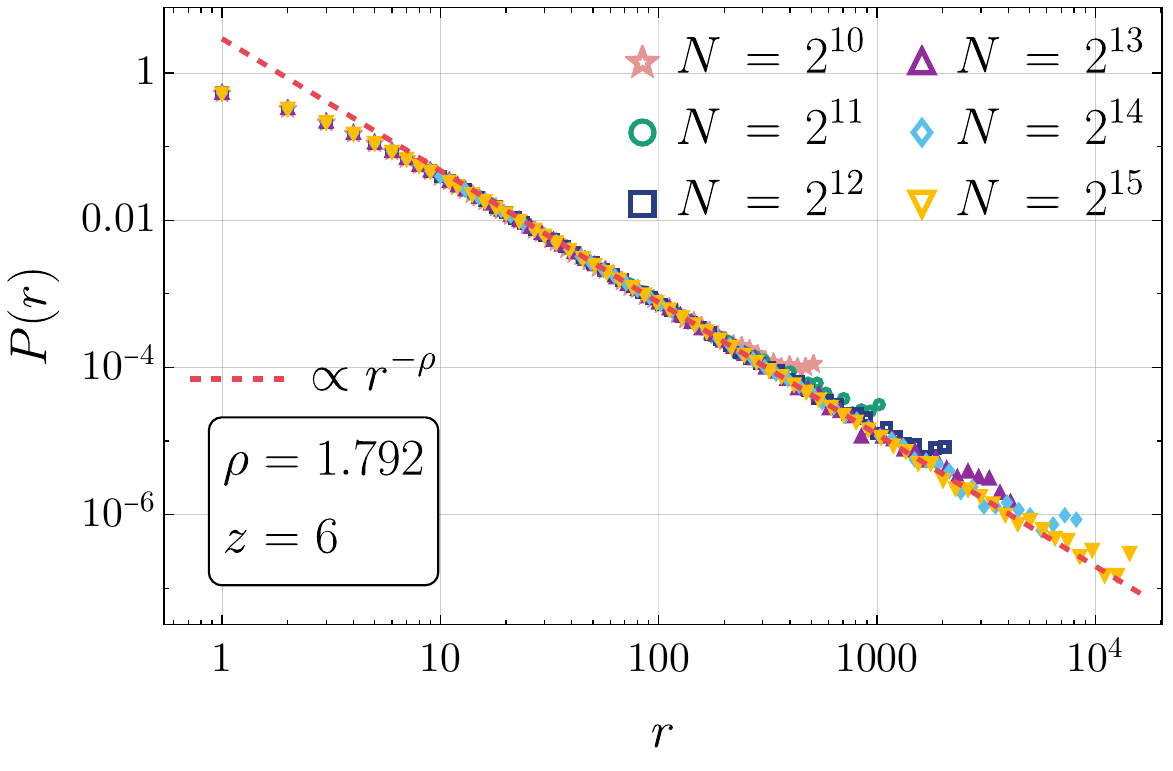}
        \caption{Numerical evaluation of the probability of edge occurrence $P(r)$. Once 1000 independent LRRGs are generated with the Monte Carlo procedure of Ref. \cite{Martin-Mayor_2012}, $P(r)$ is evaluated as the number of present edges at distance $r$ normalized by the number of possible ones. The parameters considered here are the same as the ones used in Ref. \cite{Martin-Mayor_2012}: $z=6$ and $\rho=1.792$ for different sizes $N$.}
        \label{fig:random_walk}
\end{figure}

We consider the spin glass model defined by the following Hamiltonian:
\begin{equation}
    \mathcal{H}\left(\{\sigma_i\}_{i\in \mathcal{L}_{\text{LR}}}\right)=-\sum_{( i,j)\in\mathcal{E}(G)}J_{i j}\sigma_i\sigma_j - \sum_{i\in\mathcal{L}(G)} H_i\sigma_i\,,
\end{equation}
where $H_i$ is an external magnetic field acting on $\sigma_i$, while $\mathcal{E}(G)$ and $\mathcal{L}(G)$ denote, respectively, the set of edges and set of nodes of the one-dimensional LR diluted lattice $G$, defined as follows. We consider the following set of random graphs $G_z$, characterized by fixed connectivity $z$ on the ring of length $N$. Each graph is weighted with the following factor
\begin{equation}
    w(G) = \frac{1}{Z} \prod_{(ij) \in \mathcal{E} }w_{ij}
\end{equation}
with 
\[
w_{ij} = |x_i-x_j|^{-\rho} \,,
\]
where $1<\rho<3$, $x_i$ is the coordinate of node $i$ on the ring and $Z$ is an appropriate factor that ensures the normalization of $w(G)$ over the set of graphs: $\sum_{G \in G_z} w(G)=1$.
In practice, these graphs can be generated via  Monte Carlo simulation on $G_z$ with Energy $H_{bond}(G)=-\ln w(G)$ and temperature $T=1$. For the details, we refer to Ref. \cite{Martin-Mayor_2012}. Thus, the probability $P(r)$ that two sites at distance $r=|x_i-x_j|$ are connected is not simply proportional to $w_{ij}$ but decays with the same power law at large distances.  In Fig. \ref{fig:random_walk}, we plot the probability distribution $P(r)$ as obtained from the Monte Carlo procedure. We will refer to the regular graphs, generated via this method, as Long Range Regular Graphs (LRRGs).

In order to compute the expansion in powers of $1/M$, we need to compute the average number $\mathcal{N}_L(x)$ of Non-Backtracking Paths (NBP) of length $L$ connecting two sites at distance $x$ \cite{fitzner2013non}. In particular, when dealing with models defined on finite-dimensional lattices near criticality, we are interested in the asymptotic behavior of NBP for large $L$ and $x$. This quantity is not easily accessed analytically; however, we expect that its Fourier transform, $\mathcal{N}_L(k)$, is given by
\begin{equation}
\widehat{\mathcal{N}}_L(k) \sim \exp( -|k|^{\rho-1} L) \,.
\label{NLk}
\end{equation}
More details are given in Sec. \ref{sec:random_walks}, where the behavior is verified numerically by studying NBPs on graphs generated with the algorithm of \cite{Martin-Mayor_2012}.

We are interested in  characterizing this LR model in the regime $1<\rho<3$. Indeed, for $\rho=0$ the notion of distance loses meaning, and the white measure on $G_z$ implies that the typical graph is locally tree-like. The tree-like topology remains also for $\rho \leq 1$  in the thermodynamic limit, as the nearest neighbors of each site will be at infinite distance on the chain.  This follows from the fact that $P(r) \propto r^{-\rho}$ cannot be defined for $\rho \leq 1$. For $\rho\geq 3$, the expression in Eq. (\ref{NLk}) is not valid and we have $\widehat{\mathcal{N}}_L(k) \sim \exp( -|k|^2 L) $. As a consequence, the physics becomes that of a one-dimensional model with short-range (SR) interactions, and certainly there is no phase transition.

Since we want to study a spin glass model, once the network structure is established, we assign to each existing edge a coupling constant drawn from either a Gaussian or bimodal distribution, a choice that does not affect the critical behavior. We will consider the case of zero mean and finite variance. To avoid trivial ground state degeneracy, in the case of bimodal couplings, the fields $H_i$ are i.i.d. Gaussian variables with zero mean and variance $\Delta^2$. On the other hand, for Gaussian couplings, the external field can be considered constant $H_i=H$ $\forall i=1,\dots, N$. In both cases, a zero temperature spin glass transition occurs for a finite critical value of the variance, $\Delta_c^2$, or external field, $H_c$, respectively.

Interestingly, there is a direct correspondence between the LR model and its finite-dimensional SR counterpart. In the SR case, critical behavior is governed by the spatial dimension $D$, bounded by a lower critical dimension $D_\text{lc}$ and an upper critical dimension $D_\text{uc}$. An analogous scenario holds for the one-dimensional LR model, where the exponent $\rho$ plays a role equivalent to the dimension. Here, one identifies a lower critical value $\rho_\text{lc}$ and an upper critical value $\rho_\text{uc}$. For $\rho < \rho_\text{uc}$, the system exhibits MF critical behavior, just as an SR system does for $D > D_\text{uc}$. In the intermediate regime, $\rho_\text{lc} > \rho > \rho_\text{uc}$, corrections emerge, similar to the behavior of a SR system with $D_\text{lc} < D < D_\text{uc}$; while, for $\rho > \rho_\text{lc}$, no transition is observed. The correspondence between the LR and SR models is addressed in Section \ref{subsec:Correspondence between long-range and short-range models}, where a direct relation between $D$ and $\rho$ is derived. This relation is exact above the upper critical dimension and allows us to identify $\rho_\text{uc} = 5/4$ from the known value $D_\text{uc} = 8$ for the present model. In contrast, the value of $\rho_\text{lc}$ cannot be determined analogously, as the lower critical dimension $D_\text{lc}$ for spin glasses in a field remains unknown. Moreover, the relation between $D$ and $\rho$ is only approximated below the upper critical dimension.

\subsubsection{The $M$-layer construction}
We now introduce $M$-layer construction for this model. A detailed treatment of this construction will be provided in Sec. \ref{sec:The method: M-layer construction}, where we outline the computations in detail. $M$ copies of the original graph are created, called \quotationmarks{layers}, from which the name of the construction. Inter-layer connections are generated by random rewiring of edges across different copies. As explained in Sec. \ref{sec:Description of the Method}, as $M\to\infty$, the graph becomes locally tree-like, and the critical behavior is governed by the MF fixed point. Indeed, it is possible to argue that the probability to close topological loops with the rewiring procedure is proportional to $1/M$, which serves as a small parameter to compute perturbative expansions of interesting observables, in the large-$M$ regime. Each term of the series is related to a particular topology, which we will call \quotationmarks{diagram}, with an increasing number of topological loops. For $M\to\infty$, only the first contribution is relevant, a loop less connected diagram.

Given that the rewiring is a random procedure, we want to evaluate the \textit{average} value of an observable over the procedure. This  value can then be organized as a sum over different topologies. Each term of the sum includes (i) the power of the expansion parameter, $1/M$, (ii) the observable computed on the given diagram, and (iii) the number of rewirings that realize that specific diagram. This last ingredient coincides with the number of NBP, $\mathcal{N}_L(x)$, on the chosen graph. Using the asymptotic behavior of the number of NBP at large $L$, we thus obtain the critical exponents of zero-temperature spin-glass models in a field defined on lattices that share this NBP scaling.

\subsection{Zero-temperature phenomenology}\label{subsec:Zero-temperature phenomenology}
In order to characterize the critical behavior, we first define the relevant observables computed on the original LR lattice. We begin with the notion of \textit{local field} on site $i$, denoted $h_i$, defined as the effective field acting on spin $\sigma_i$ in the ground state configuration. By definition, $\sigma_i$ aligns with the sign of $h_i$ in the ground state, and the \emph{excitation energy}, $\Delta E_i = 2|h_i|$, corresponds to the energy required to flip $\sigma_i$. In other words, $\Delta E_i$ is the energy difference between the ground state and the configuration in which $\sigma_i$ is constrained to point opposite to $h_i$. The local field $h_i$ on a given site at position $x_1$ is a non-trivial function of both $J_{ij}$ and $H_i$ $\forall i,j\in\mathcal{L}_{LR}$, which are continuous, random parameters. We therefore focus on the probability density distribution of $h_i$ taking the value $h$, denoted by $P_1(x_1,h)$. Averaging over disorder realizations, understood as both the specific graph realization and the sampling of couplings and fields, one finds that the distribution is independent of position: $P_1(x_1,h)=P_1(h)$. Given two sites $\sigma_i$ and $\sigma_j$, the notion of local field can be generalized to the triplet of effective parameters $u_i$, $u_j$ and $J_{ij}^\text{eff}$, corresponding respectively to effective fields and an effective coupling:
\begin{equation}
    E_{ij}(\sigma_i,\sigma_j)=-u_i \sigma_i-u_j \sigma_j - J_{ij}^\text{eff} \, \sigma_i \sigma_j\,,
\end{equation}
where $E_{ij}(\sigma_i,\sigma_j)$ represents, up to an irrelevant constant, the ground-state energy of the system as a function of the configurations of the two spins. As in the case of the local field, the triplet is given by a non-trivial function of the random couplings and external fields; accordingly, we define the probability density that the triplet $(u_i,u_j,J_{ij}^\text{eff})$ takes the value $(u,u',J)$ as $P_2(x_1,x_2; u,u',J)$, where $x_1$ and $x_2$ denote the lattice positions of sites $i$ and $j$, respectively. In addition to translational invariance, the distribution is symmetric with respect to the exchange $u \leftrightarrow u'$. 
One can see that, at large distances $|x_1-x_2|$, this distribution converges to the factorized form $P_1(u)P_1(u')\delta(J)$ with $P_1(u)$ the probability distribution of the local field on a given site \cite{angelini2025criticalexponentsspinglass}. Moreover, the triplet distribution is such that, at any distance $|x_1-x_2|>1$, there is a {\it finite} probability of having $J=0$, therefore we can define two additional distributions $P^\text{dis}(x_1,x_2;u,u')$ and  $P^\text{con}(x_1,x_2;u,u',J)$ according to:
\begin{multline}
P_2(x_1,x_2;u,u',J)-P_1(u)P_1(u')\,\delta(J) =\\= P^\text{dis}(x_1,x_2;u,u')\,\delta (J)+ P^\text{con}(x_1,x_2;u,u',J) ,  
\label{eq:Ptriplet}
\end{multline}
where $P^\text{con}(x_1,x_2;u,u',J)$ is regular at $J=0$.

The critical behavior can be rephrased in terms of clusters of spins with diverging size that share the same excitation energy $\Delta E_i = 2,|h_i|$. We therefore define the generic \quotationmarks{$q$-point} correlation function $\mathcal{P}_q(x_1,\dots,x_q;\Delta E)$ as the probability density that the $q$ spins at positions $x_1,\dots,x_q$ belong to the same cluster with excitation energy $\Delta E$. For $q=2$, this probability is related to the non-zero coupling contribution to $P_2(x_1,x_2;u,u',J)$, namely $P^\text{con}(x_1,x_2;u,u',J)$. Indeed, one expects that two spins located at positions $x_1$ and $x_2$ belong to the same cluster if the absolute value of the effective coupling between them is larger than both the absolute values of the effective fields $u$ and $u'$. More generally, for a given excitation energy $\Delta E$, one can identify clusters of spins that flip collectively. At zero temperature, the critical clusters are those with excitation energy $\Delta E=0$, which constitute the regime of interest in this work. In the following, we refer to these clusters as \quotationmarks{soft}.

Motivated by the percolation problem, we also consider the statistics of cluster sizes. We define the cluster density $n(s,\Delta E)$ such that, in a system of size $N$, the number of clusters of size $s$ and excitation energy between $\Delta E$ and $\Delta E+\mathrm{d}E$ is given by $n(s,\Delta E) N \mathrm{d}E$. In principle, $n(s,\Delta E)$ depends on the given realization of the disorder, but on general grounds, we expect it to be self-averaging at large $N$.
Much as in percolation, one sees that the correlation functions are related to the moments of $n(s,\Delta E)$; in particular, we have 
\begin{equation}
  \sum_s  \, s \, n(s,\Delta E) = \mathcal{P}_1(x_1;\Delta E) \ ,
\end{equation}
where we recall that $\mathcal{P}_1$ is the probability distribution that a node belongs to a cluster of excitation energy $\Delta E$ and is related to $P_1$ through $\mathcal{P}_1(x_1;\Delta E)=P_1(\Delta E/2)/2+P_1(-\Delta E/2)/2$. Furthermore, the sum of $ \mathcal{P}_{2}(x_1,x_2;\Delta E)$ over $x_2$ is related to the second moment of the cluster distribution:
\begin{equation}
  \sum_s  \, s^2 \, n(s,\Delta E) = \sum_{x_2} \mathcal{P}_{2}(x_1,x_2;\Delta E) \, 
\label{eq:s2}
\end{equation}
and in full generality:
\begin{equation}
  \sum_s  \, s^q \, n(s,\Delta E) = \sum_{x_2,\dots,x_q} \mathcal{P}_q(x_1,\dots,x_q;\Delta E) \, . 
\label{eq:s^qP_q}
\end{equation}
Note that the RHS of the last three formulas does not depend on $x_1$ due to translational invariance.

The characterization of the zero-temperature critical behavior also requires the definition of an additional observable \cite{Bray_1985}, namely the two-point disconnected correlation function $P^\text{dis}(x_1,x_2;0,0)$. Within this framework, this function describes correlations between two spins belonging to distinct soft clusters. Again, by translational invariance, both $\mathcal{P}_q$ and $P^\text{dis}$ only depend on the distances between the involved positions. For the following computations, the relevant observables to be computed on the $M$-layer graph are the susceptibilities, defined as
\begin{equation}\label{eq:chiqdef}
    \chi_q\equiv\sum_{x_2,\dots,x_q} \mathcal{P}_q(x_1,\dots,x_q;0)\,,
\end{equation}
and
\begin{equation}\label{eq:chidisdef}
    \chi_2^\text{dis}\equiv\sum_{x_2} P^\text{dis}(x_1,x_2;0,0)\,.
\end{equation}

\subsection{Scaling laws}\label{subsec:Scaling laws}
At leading order in the $1/M$ expansion, discussed in Sec. \ref{sec:computation of the susceptibilites}, Eq. \eqref{eq:Ptriplet} takes the following simple form after Fourier transform ($FT$) in position space: 
\begin{widetext}
\begin{multline}
   FT\left[ P_2(x,x';u,u',J)-P_1(u)P_1(u')\delta(J)\right] \propto 2 \pi \delta(k+k') \widetilde{P}_1(u) \widetilde{P}_1(u')\left(- \frac{1}{(k^{\rho-1}+t)^2} \delta(J) + \frac{1}{(k^{\rho-1}+t+|J|)^3}\right)\,,
\label{eq:FTP}
\end{multline}
\end{widetext}
where $t$ vanishes linearly at the critical point, i.e. $t \propto H-H_c$ or  $t \propto \Delta-\Delta_c$. The function $\widetilde{P}_1(u)$ depends on the microscopic details of the model and satisfies i) $\widetilde{P}_1(u) \geq 0$ for all $u$ and ii) $\widetilde{P}_1(u)=\widetilde{P}_1(-u)$ in the case of bimodal couplings and random Gaussian fields \footnote{Note that, applying a random transformation $\sigma_i \to -\sigma_i$ with probability $1/2$ independently for each spin, the constant field case reduces to the random field case with $H_i = \pm H$.}. As a consequence of $\widetilde{P}_1(u)=\widetilde{P}_1(-u)$, the average over the disorder of $\sigma_i \sigma_j$  on the ground state vanishes. 

Moreover, given a generic translationally invariant two-point function $C_2(x,x')$, one can define the associated correlation length on the one-dimensional LR lattice as
\begin{equation}\label{eq:def_corr_length}
    \xi^{\rho-1}\equiv\left(\frac{\mathrm{d}}{\mathrm{d}k^{\rho-1}}\widehat{C}_2(k)^{-1}\right)\Bigg|_{k=0}\widehat{C}_2(0)\,,
\end{equation}
where $\widehat{C}_2(k)=FT[C_2(x,x')]$. The physical interpretation of the correlation length for LR model is given in \ref{subsec:Double power law} (see also Ref. \cite{joyce1966spherical}). Thus in Eq. \eqref{eq:FTP} 
two correlation lengths $\xi_\text{dis}$ and $\xi_\text{con}$ can be identified, respectively associated with the disconnected (the $J=0$ term) and connected (the $J\neq0$ term) correlation functions.  
The length $\xi_\text{dis}$ diverges as $t^{-1/(\rho-1)}$, while $\xi_\text{con} \sim (t+|J|)^{-1/(\rho-1)}$ and it depends on both $t$ and $J$ (but not on $u$ and $u'$) and diverges iff both $t$ and $J$ vanish.
We note that Eq. \eqref{eq:FTP} holds at small momenta, that is at large distances $|x-x'|$, of the order of the correlation lengths that are large close to the critical point $(t=0)$ when $|J|\ll 1$. 

As mentioned above, higher-order corrections in $1/M$ do not modify Eq. \eqref{eq:FTP} below the upper critical value $\rho_\text{uc}=5/4$. For $\rho>5/4$, such corrections become relevant; nevertheless, we expect Eq. \eqref{eq:FTP} to be generalized by the following scaling form, characterized by three critical exponents $(\theta,\nu,\overline{\eta})$:
\begin{widetext}
\begin{multline}
    P_2(x_1,x_2;u,u',J)-P_1(u)P_1(u')\delta(J)=\widetilde{P}_1(u) \widetilde{P}_1(u') \left(- \frac{1}{r^{3-2\rho+\overline{\eta}}} f_\text{dis}\left( \frac{r}{\xi_\text{dis}}\right) \delta(J) +\frac{1}{r^{3-2\rho+\overline{\eta} -\theta}} f_\text{con}\left( \frac{r}{\xi_\text{con}}, \frac{|J|}{t^{\nu \theta}}\right) \right)\,,
\label{eq:Pscaling}
\end{multline}
\end{widetext}
where $r \equiv |x_1-x_2|$. The length $\xi_\text{dis}$ depends only on $t$ and diverges as $\xi_\text{dis} \propto t^{-\nu}$, while $\xi_\text{con}$ depends on both $t$ and $J$ (but not on $u$ and $u'$) and obeys the scaling form $\xi_\text{con}=t^{-\nu}f_{\xi}(|J|/t^{\nu \theta})$.
The scaling function $f_{\xi}$ is such that $\xi_\text{con}$ remains finite unless both $t$ and $J$ vanish. In particular, $\xi_\text{con} \propto t^{-\nu}$ for $J=0$ and $\xi_\text{con} \propto |J|^{-1/\theta}$ for $t=0$.
For $\rho>\rho_\text{uc}$, the critical exponents $(\theta,\nu,\overline{\eta})$ and the scaling functions $f_\text{dis}(\varrho)$, $f_\text{con}(\varrho,\jmath)$ and $f_{\xi}(\jmath)$ depend on $\rho$ but not on the microscopic details of the model, and are therefore universal. The scaling variable $\jmath$ in $f_\text{con}(\varrho,\jmath)$ and $f_{\xi}(\jmath)$ parametrizes the lines of approach to the point $(t=0,J=0)$, ranging from $\jmath=0$, corresponding to $J=0$, to $\jmath=\infty$, corresponding to the line $t=0$.
Note that Eq. \eqref{eq:FTP} is recovered as a special case of Eq. \eqref{eq:Pscaling} with $\theta=\rho-1$, $\nu=1/(\rho-1)$, $\overline{\eta}=0$, and suitable choices of the scaling functions $f_\text{con}(\varrho,\jmath)$, $f_\text{dis}(\varrho)$ and $f_{\xi}(\jmath)$. 

As in Eq. \eqref{eq:FTP}, the function $\widetilde{P}_1(u)$ is positive definite and model-dependent, in contrast with the universal functions $f_\text{dis}$ and $f_\text{con}$. Since, by definition, i) $P_2(x_1,x_2;u,u',J)$ is normalized with respect to $(u,u',J)$ and ii) $P_1(u)$ is normalized with respect to $u$, it follows that the integral over $(u,u',J)$ of the LHS of Eq. \eqref{eq:Pscaling} vanishes. On the other hand, since $\widetilde{P}_1(u)$ on the RHS is positive definite, the vanishing of the RHS must originate from the integration over $J$. This leads to: i) the same exponent $\overline{\eta}$ appears in the connected and disconnected part, ii) $f_\text{dis}(\varrho)$ is related to $f_\text{con}(\varrho,\jmath)$ and $f_{\xi}(\jmath)$ by the following relationship: 
\begin{equation}
f_\text{dis}\left(\frac{r}{\xi_\text{dis}(t)}\right)=\int_{-\infty}^{\infty} f_\text{con}\left( \frac{r}{\xi_\text{con}(t,J)}, \frac{|J|}{t^{\nu \theta}}\right)\,\mathrm{d}J  \ ,
\label{cond1}
\end{equation}
iii) the ratio $\xi_\text{dis}(t)/\xi_\text{con}(t,J)$ depends solely on the ratio $|J|/t^{\nu \theta}$ and is universal.

As discussed in Sec. \ref{subsec:Zero-temperature phenomenology}, the connected correlation function $\mathcal{P}_2(x_1,x_2;\Delta E)$ can be obtained from the distribution $P_2(x_1,x_2;u,u',J)$ by integrating $P_2(x_1,x_2;u,u',J)$ over $u$, $u'$ and $J$ under the constraints $|u|<|J|$, $|u'|<|J|$, and $|u+u'\,\sign J|=\Delta E/2$. 
A straightforward calculation then yields, from Eq. \eqref{eq:FTP}, the following expression valid for $\rho \leq \rho_\text{uc}$:
\begin{equation}
   FT[\mathcal{P}_2(x_1,x_2;\Delta E)] \propto
    \frac{2 \pi \delta(k+k')
    \widetilde{P}_1^2(0)}{k^{\rho-1}+t+\Delta E/4} \ .
\label{eqetaMF}
\end{equation}
The associated correlation length is $\xi=(t+\Delta E/4)^{-1/(\rho-1)}$ and therefore diverges only for $\Delta E=0$ and $t=0$, implying that only soft clusters display critical behavior. For $\rho > \rho_\text{uc}$, this generalizes to the following scaling form, valid at large $r$
\begin{equation}
    \mathcal{P}_2(x_1,x_2;\Delta E) =
    \widetilde{P}_1^2(0)\, \frac{1}{r^{3-2\rho+\overline{\eta}+\theta}}  \, f_2\left( \frac{r}{\xi}, \frac{\Delta E}{t^{\nu \, \theta}}\right)
\label{eqeta2}
\end{equation}
where $f_2(r/\xi,\Delta E/t^{\nu\theta})$ is a scaling function, related to $f_\text{con}$ and $f_{\xi}$. The correlation length $\xi$ depends on both $t$ and $\Delta E$; similarly to $\xi_\text{con}$, it obeys the scaling form $\xi=t^{-\nu}\tilde{f}_\xi(\Delta E/t^{\nu \theta})$ and diverges as $\xi \propto t^{-\nu}$ for $\Delta E=0$ and as $\xi \propto \Delta E^{-1/\theta}$ for $t=0$ \footnote{Note that the universal scaling function of $\xi_\text{con}(t,J)$ is not the same as that of $\xi(t,\Delta E)$.}. 
Defining the exponent $\eta$ from $\mathcal{P}_2(x_1,x_2;\Delta E) \propto 1/r^{2-\rho+\eta}$ we have 
\begin{equation}
    \eta= \overline{\eta}-\rho+1+\theta \ 
\label{eq:eta}
\end{equation}
and for $\rho \leq 5/4$ the values $\theta=\rho-1$, $\overline{\eta}=0$  imply the usual result $\eta=0$.

\subsubsection{The double power law}\label{subsec:Double power law}
For SR models, the two-point connected correlation function exhibits a power law scaling for $r/\xi\to0$ and decreases exponentially for $r/\xi\to\infty$; instead, for LR models, there is an important difference, indeed it shows a double power law, with different exponents \cite{behan2017scaling}. Indeed, considering Eq. \eqref{eqeta2} with $\Delta E=0$, that is the two-point connected correlation function relevant at zero temperature, one can write
\begin{equation}\label{eq:scaling_2point}
    \mathcal{P}_2(0,r;0) \propto  \int_{-\infty}^\infty  \frac{e^{ikr}}{|k|^{\rho-1}+\xi^{-\rho+1}}\mathrm{d}k\,,
\end{equation}
that is valid in the MF approximation (leading $1/M$ term in the $M$-layer construction). Rescaling the momentum by $k=u/r$, we can write the scaling function $f_2$ as defined in Eq. \eqref{eqeta2}, in the MF regime
\begin{equation}
f_2(z,0)\equiv\int_0^{\infty} \frac{\cos\left(u\right)}{u^{\rho-1}+z^{\rho-1}}\mathrm{d}u\,.
\end{equation}
From this scaling form one obtains the two asymptotic regimes: $r\ll\xi$ and $r\gg\xi$. For $z=r/\xi\ll1$, the relevant contribution of the integral in $f_2(z,0)$ is given by large $u$, such that, for $1<\rho<2$ 
\begin{multline}
    f_2(z\ll1,0)\simeq \int_0^\infty \cos(u)\,u^{1-\rho}\mathrm{d}u = \\\sin\left(\frac{\pi(\rho-1)}{2}\right)\Gamma\left(2-\rho\right)\,.
\end{multline}
Substituting this evaluation in Eq. \eqref{eqeta2}:
\begin{equation}\label{eq:scaling_2point_criticality}
\mathcal{P}_2(0,r;0)\sim  r^{\rho-2} \qquad \text{for}\quad r/\xi\ll1.
\end{equation}
On the other hand, for $z=r/\xi\gg1$, the integral can be rewritten with integration by parts as follows
\begin{multline}
f_2(z\gg1,0)=(\rho-1)\int_0^{\infty} \frac{\sin(u)\,u^{\rho-2}}{\left(z^{\rho-1}+u^{\rho-1}\right)^2}\mathrm{d}u\simeq\\
(\rho-1)\,z^{-2(\rho-1)}\int_0^{\infty} \sin(u)\,u^{\rho-2}\mathrm{d}u = \\z^{-2(\rho-1)}\,\Gamma\left(\rho\right)\sin\left(\frac{\pi(\rho-1)}{2}\right)\,,
\end{multline}
again for $1<\rho<2$ and where the approximation in the denominator is justified if $z\gg1$. The resulting correlation function reads
\begin{equation}\label{eq:scaling_2point_notcritical}
\mathcal{P}_2(0,r;0)\sim  r^{-\rho} \qquad \text{for}\quad r/\xi\gg1.
\end{equation}
Thus, the propagator exhibits two power-law regimes: it decays as $r^{\rho-2}$ at criticality and as $r^{-\rho}$ away from criticality. Near $\rho_\text{uc}=5/4$ (from below, where the MF approximation applies), the two-point connected susceptibility diverges, since the power $r^{\rho-2}$ is not integrable at large $r$. On the other hand, near the critical point, there is a cut-off for $r>\xi$, and the power-law becomes $r^\rho$, which, in turn, is integrable. Indeed, in these LR systems, the correlation length $\xi$ assumes the meaning of crossover length scale between the two power law regimes, at variance with SR ones, where it is defined as the scale of exponential decay of the two-point correlation function.

Moreover, here we are studying the MF regime, $\rho<5/4$, but we expect the double power-law to hold even above $\rho_\text{uc}=5/4$. The previous scaling analysis is not consistent with the values $\rho\ge2$, for which we do not expect to observe a phase transition, according to previous results in Refs. \cite{Leuzzi_2009,Martin-Mayor_2012,vedula2023study}, that find $\rho_\text{lc}<2$.

Similarly, one can observe analogous power-law regimes for the disconnected function, which can be written in the following scaling form:
\begin{multline}
    P_2^\text{dis}(0,r;0,0)\propto \\\int_{-\infty}^{\infty}  \frac{e^{ikr}}{(|k|^{\rho-1}+\xi_\text{dis}^{-\rho+1})^2} \mathrm{d}k = \frac{1}{r^{1-2(\rho-1)+\overline{\eta}}}f_\text{dis}\left(r/\xi_\text{dis}\right)\,.
\end{multline}

\subsubsection{Scaling of susceptibilities}

Given the scaling forms of the correlation functions, we can obtain the corresponding form of the susceptibilities defined in Eqs. \eqref{eq:chiqdef} and \eqref{eq:chidisdef}. Approaching the critical point $(t=0,\Delta E=0)$, the space integral of Eq. \eqref{eqeta2} diverges with $\xi$  as 
\begin{equation}
 \sum_{x_2}  \mathcal{P}_{2}(x_1,x_2;\Delta E) \,  \propto \xi^{\rho-1-\eta}  \, \tilde{f}_2\left( \frac{\Delta E}{t^{\nu\,\theta}}\right)
\label{eq:X2_scal}
\end{equation}
where $\tilde{f}_2(e)$ is related to $f_2(\varrho,e)$  (see the Supporting Information of Ref. \cite{angelini2025criticalexponentsspinglass}).
The corresponding expression for $\rho \leq \rho_\text{uc}$ is given by Eq. \eqref{eqetaMF} evaluated for $k=0$ leading to
\begin{equation}
 \sum_{x_2}  \mathcal{P}_{2}(x_1,x_2;\Delta E) \,  \propto \frac{1}{t+\Delta E/4} \propto \xi^{\rho-1}\,,
\end{equation}
where we used $\eta=0$.

To obtain the scaling form of the generic $q$-point susceptibility, the scaling form of the cluster number density is required. Following the analogous treatment of percolation \cite{coniglio2000geometrical,angelini2025bethe}, we assume that, changing the length-scale by a factor $b$ by means of a real-space RG transformation, the size $s$ of large clusters, the value of the typical energy excitation energy $\Delta E$, and the reduced field $t$ change according to:
\begin{equation} 
  s'=s/b^{D_f} , \ \Delta E' = \Delta E/b^{-\theta}, \ t'=t/b^{-1/\nu}, \ \xi' = \xi/b\ .  
\end{equation}
$D_f$ is by definition the fractal dimension of the clusters: it relates the linear size $l$ of a large cluster with its size $s$ using $s \propto l^{D_f}$. Now we {\it assume} that under the RG transformation the cluster number at large $s$ is conserved, i.e.
\begin{equation}
   n(s,\Delta E,t) \Delta s \, \Delta (\Delta E) L \approx  n(s',\Delta E',t') \Delta s' \, \Delta  (\Delta E') L'
\label{hyphyp}
\end{equation}
where $L$ is the size of the system. 
Setting $b=(s/S)^{1/D_f}$ for some fixed large reference size $S$ we then obtain :
\begin{equation}
    n(s,\Delta E,t) \propto s^{\frac{\theta-1-D_f}{D_f}} n\left(S, \frac{\Delta E}{ (s/S)^{-\theta/D_f}},\frac{t}{(s/S)^{-1/(\nu D_f)}}\right)
\end{equation}
that can be rewritten in terms of $\xi(t,\Delta E)$ (the correlation length of $\mathcal{P}_{2}(x_1,x_2;\Delta E)$) as 
\begin{equation}
    n(s,\Delta E,t) \equiv s^{(\theta-1-D_f)/D_f} f_0(s / \xi^{D_f}, \Delta E / t^{\nu \theta})\ .
    \label{nlowd}
\end{equation}
$f_0(s,e)$ is a scaling expression for large values of $s$ and has a finite limit for $s=0$. 
Therefore, the cluster density at fixed $\Delta E$ and $t$ follows a power-law $s^{(\theta-1-D_f)/D_f}$ up to a large value $s^*$ that diverges as $\xi^{D_f}$ approaching the critical point $t=0$ when $\Delta E\sim 0$.
We remark that the clusters with a finite $\Delta E>0$ are characterized by a finite correlation length, including at $t=0$. Indeed we have $s^* \sim t^{-\nu \, D_f}$ at $\Delta E=0$ and $s^* \sim \Delta E^{-D_f/\theta}$ at $t=0$.
From Eq. \eqref{eq:s^qP_q} and Eq. \eqref{nlowd} we obtain:
\begin{multline}
 \sum_{x_2,\dots,x_q} \mathcal{P}_q(x_1,\dots,x_q;\Delta E) = \\ \sum_s  \, s^q \, n(s,\Delta E,t)  \equiv \xi^{D_f q-1 + \theta}\tilde{f}_{q}(\Delta E/t^{\nu \theta})\ . 
 \label{eq:def_K-point_function}
\end{multline}
In the following, we define the susceptibilities as the above integrals for $\Delta E=0$. They diverge at the critical point as:
\begin{equation}
    \chi_q \equiv \!\!\sum_{x_2,\dots,x_q}\!\! \!\!\mathcal{P}_q(x_1,\dots,x_q;0) \propto \xi^{D_f q -1+ \theta} \propto t^{-\nu (D_f q-1 + \theta)}\ .
    \label{eq:def_K-point_susceptibility}
\end{equation}
The comparison of Eq. \eqref{eq:def_K-point_function} for $q=2$ with Eq. \eqref{eq:X2_scal} leads to the identification of the energy exponent with $\theta$ and to:
\begin{equation}
    D_f=\frac{\rho-\theta-\eta}{2}\, ,\ \ \chi_q \propto \xi^{\left(\frac{q}{2}-1\right)(1-\theta)+\frac{q}{2}(\rho-1-\eta)}\ .
\label{eq:X2scalb}
\end{equation}
Note that this leads to $D_f=1/2$ for $\rho=\rho_\text{uc}$.
Analogously to the connected susceptibility $\chi_2$, we introduce the disconnected susceptibility as:
\begin{multline}
   \chi_2^\text{dis} \equiv \sum_{x_2} P^\text{dis}(x_1,x_2;0,0) \\ \propto -  \int_0^\infty \frac{1}{r^{3-2\rho+\overline{\eta}}} \,f_\text{dis}\left( \frac{r}{\xi}\right)  \mathrm{d}r \propto -\xi^{2(\rho-1)-\overline{\eta}}  \ .
   \label{eqeta1}
\end{multline}

\subsection{Correspondence between long-range and short-range models}\label{subsec:Correspondence between long-range and short-range models}
Remarkably, one can identify a connection between the decay exponent $\rho$ for the LR model and the physical dimension $D$ of the corresponding SR version \cite{larson2010numerical,Martin-Mayor_2012}. This correspondence is exact in the MF regime, whereas it is only an approximation below the upper critical dimension. We repeat here the same argument of Ref. \cite{Martin-Mayor_2012}, generalizing it to the zero-temperature case in which the additional exponent $\theta$ is present. We make use of the scaling form of the singular part of the cluster density of a generic system in $D$ dimensions \cite{angelini2025criticalexponentsspinglass}
\begin{equation}
    n_\text{sing}(s,\Delta E,t)\! =\! \frac{1}{L^{D-\theta+D_f}} f_\text{sing}(L^{y_s}s, L^{y_e}\Delta E, L^{y_t}t, L^{y_u}u ),
\end{equation}
where $t$ is the distance from the critical point, $u$ is the operator which gives the leading correction to scaling, and $f_\text{sing}$ is a scaling function. The exponents $y_s,\,y_e,\,y_t$ and  $y_u,$ can be written in term of the ones previously defined
\begin{equation}
    y_s = -D_f\,,\quad y_e=\theta\,,\quad y_t=\frac{1}{\nu}\,,\quad y_u=\omega\,, 
\end{equation}
where $\omega<0$ is the correction-to-scaling critical exponent. Now we compare the cluster densities of the SR with those of the (one-dimensional) LR, with the same number of spins $L_\text{sr}^D=L_\text{lr}$ and with the same couplings variance $J_\text{sr}=J_\text{lr}$:
\begin{multline}\label{eq:cluster_densities_comparison}
    \frac{1}{L_\text{sr}^{D-y^\text{sr}_e-y^\text{sr}_s}}\,f_\text{sr}(L_\text{sr}^{y_s^\text{sr}}s,L_\text{sr}^{y_e^\text{sr}}\Delta E,L_\text{sr}^{y_t^\text{sr}}t,L_\text{sr}^{y_u^\text{sr}}u)=\\ \frac{1}{L_\text{lr}^{1-y^\text{lr}_e-y^\text{lr}_s}}\,f_\text{lr}(L_\text{lr}^{y_s^\text{lr}}s,L_\text{lr}^{y_e^\text{lr}}\Delta E,L_\text{lr}^{y_t^\text{lr}}t,L_\text{lr}^{y_u^\text{lr}}u)\,.
\end{multline}
Using $L_\text{sr}^D=L_\text{lr}$ and imposing the validity of Eq. \eqref{eq:cluster_densities_comparison} we obtain:
\begin{equation}\label{eq:correspondence_SR-LR}
    \frac{y_i^\text{sr}}{D}=y_i^\text{lr} \,,\quad \text{for } i=s,\,t,\,e,\, u\,.
\end{equation}
Particularly relevant is the relation $y_s^\text{sr}/D=y_s^\text{lr}$ that can be written using the definition of $D^{\text{lr}}_f$ in Eq. \eqref{eq:X2scalb} (and the corresponding SR one, $D^\text{sr}_f=\frac{D-\theta_\text{sr}+2-\eta_\text{sr}}{2D}$, see Ref. \cite{angelini2025criticalexponentsspinglass}) considering $\eta_\text{lr}=0$
\begin{equation}
    \frac{D-\theta_\text{sr}+2-\eta_\text{sr}}{2D} = \frac{\rho-\theta_\text{lr}}{2}
\end{equation}
from which we obtain the relation between $\rho$ and the physical dimension $D$
\begin{equation}\label{eq:Dim_SR_LR_correspondence}
    D=\frac{2-\eta_\text{sr}}{\rho-1}\,.
\end{equation}

\subsection{Mean-field behavior}
Here we describe the behavior of the model defined on the BL with connectivity $z$. This choice is motivated by the fact that, in the limit $M\to\infty$, the $M$-layer construction applied to an instance of the LRRG yields precisely this graph. The first tool we need is the cavity distribution of the \quotationmarks{bias} $u$
\begin{multline}
    P_B(u)=\mathbb{E}_{J,H}\,\int\,\prod_{i=1}^{z-1}P_B(u_i)\,\mathrm{d}u_i\times\\\times\delta\!\Bigg( \!u - \sign\Big( J\big(H+\sum_{i=1}^{z-1} u_i\big) \Big)\! \min\!\Big( |J|,|H+\sum_{i=1}^{z-1}u_i| \Big)\!\Bigg)\,,
\end{multline}
where the average over disorder, encoded in couplings and/or external fields, is denoted with $\mathbb{E}_{J,H}$ and $H$ is the fixed external field (in the case of Gaussian couplings) or an instance of the Gaussian external field (in the case of bimodal couplings). The distribution of the fields $P_1(h)$ on the BL is then given by:
\begin{equation}
    P_1(h)=\mathbb{E}_{H}\int \prod_{j=1}^zP_B(v_j)\mathrm{d}v_j\,\delta\Big(h-(H+\sum_{j=1}^zv_j)\Big)\,.
\end{equation}
We now want to characterize the distribution of the triplets, as introduced in Sec. \ref{subsec:Zero-temperature phenomenology}, between two spins of connectivity 1 that are placed at a large distance $L$ on a BL. In this case, in order to characterize the critical behavior, we can make use of the following \textit{Ansatz}, that we call $P_L(u_1,u_2,J)$:
\begin{multline}\label{eq:Ansatz}
   \hspace{-0.4cm} P_L(u_1,u_2,J)\equiv \delta(J)P_B(u_1)\,P_B(u_2)\\-2aL\widetilde{\lambda}^L\delta(J)\,g(u_1)g(u_2)+aL^2\widetilde{\lambda}^L\widetilde{\rho}e^{-\widetilde{\rho}|J|L}g(u_1)g(u_2),
\end{multline}
where $a$ and $\widetilde{\rho}$ (not to be confused with $\rho$, the exponent of the LR interaction) are computable parameters, $\widetilde{\lambda}<1$ and $g(u)$ are respectively the largest eigenvalue and the corresponding eigenfunction of a specific integral equation \cite{Parisi_2014}
\begin{multline}
  \widetilde{\lambda}\,g(u)=\mathbb{E}_J\int\prod_{i=1}^{z-2}P_B(u_i)\mathrm{d}u_i\,g(u')\mathrm{d}u'\times \\\times \Theta\Big(|J|-|H+u'+\sum_{i=1}^{z-2}u_i|\Big)\delta\Bigg( u-\sign(J)|H+u'+\sum_{i=1}^{z-2}u_i| \Bigg),
\end{multline}
where $\Theta(x)$ is the Heaviside step function. The control parameter is identified in $\widetilde{\lambda}$, which takes the value $1/(z-1)$ at the MF critical point.

We remark that the \textit{Ansatz} given in Eq. \eqref{eq:Ansatz}, is an approximation for the distribution of the triplet on a chain of length $L$, in which the external spins have connectivity equal to 1 and the internal ones equal to $z$. This construction allows to join lines as follows: if we want to join two lines characterized by two triplets $(u_1,u_2,J)$ and $(u_2', u_3,J')$ on a central site, the resulting local field on this central site will be the sum of the two biases $u_2$ and $u_2'$ coming from the two lines, and then of $z-2$ more fields. Generalizing this procedure to attach $n$ lines in a single site (which we also call \quotationmarks{$n$-degree vertex}), the field on the central spin will thus be distributed according to $\widetilde{P}_n(h)$:
\begin{multline}
    \widetilde{P}_n(h)\equiv \mathbb{E}_H\int\!\prod_{i=1}^{n}g(u_i)\mathrm{d}u_i\times \\ \times\prod_{j=n+1}^{z}\!\!\!P_B(u_j)\mathrm{d}u_j\,\delta\Big(h-H-\!\sum_{i=1}^{n}\!u_i-\!\sum_{j=n+1}^{z}\!u_j\Big)\,.
\end{multline}
In order for the \textit{Ansatz} to be \quotationmarks{self-consistent} (that is its form for a chain of length $L_1+L_2$ should be the same as the one resulting from the connection of two lines, with lengths $L_1$ and $L_2$) $\widetilde{P}_2(h)$ satisfies the following relation,
\begin{equation}
    \frac{4\,a\,\widetilde{P}_2(0)}{\widetilde{\rho}}=1\,.
\end{equation}
While the choice of assuming the connectivity of the external spins in Eq. \eqref{eq:Ansatz} equal to 1 is very useful when one wants to join lines together, in order to compute observables, we will need to insert $z-1$ additional fields, on the external spins in such a way that they result to be typical spins extracted from a BL. In this way the \textit{Ansatz} becomes 
\begin{multline}\label{eq:modified_Ansatz_expression}
   Q_L(h_1,h_2,J)= \delta(J)P_1(h_1)P_1(h_2) \\-2aL\widetilde{\lambda}^L\delta(J)\widetilde{P}_1(h_1)\widetilde{P}_1(h_2)\\+aL^2\widetilde{\lambda}^L\widetilde{\rho}e^{-\widetilde{\rho}|J|L}\widetilde{P}_1(h_1)\widetilde{P}_1(h_2)
\end{multline}
that corresponds to Eq. \eqref{eq:Ptriplet}, but written here for the BL.

\begin{figure}[h]
        \centering        \includegraphics[scale=0.3]{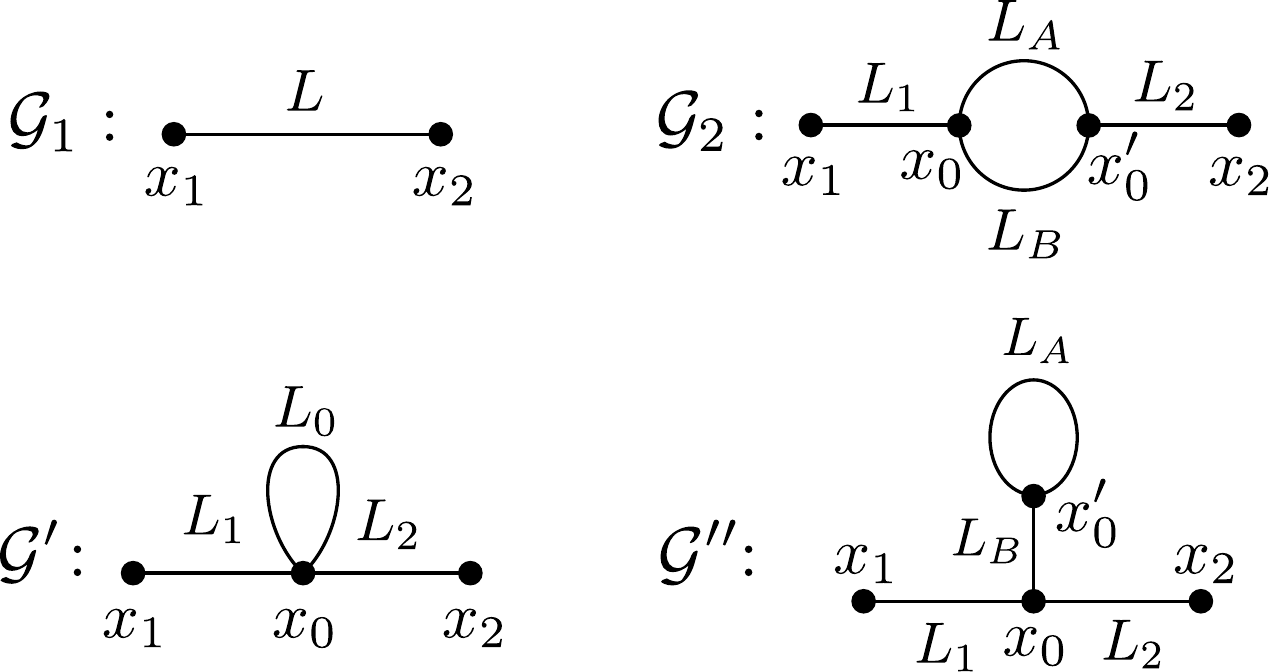}
        \caption{Diagrams considered for the evaluation of two-point correlation functions (both connected and disconnected) up to one-loop order. Diagram $\mathcal{G}_1$ contributes to $\mathcal{O}(1/M)$ while diagrams $\mathcal{G}_2$, $\mathcal{G}'$ and $\mathcal{G}''$ contribute to $\mathcal{O}(1/M^2)$.}
        \label{fig:diag_CD2}
\end{figure}

\begin{figure}[h]
        \centering        \includegraphics[scale=0.28]{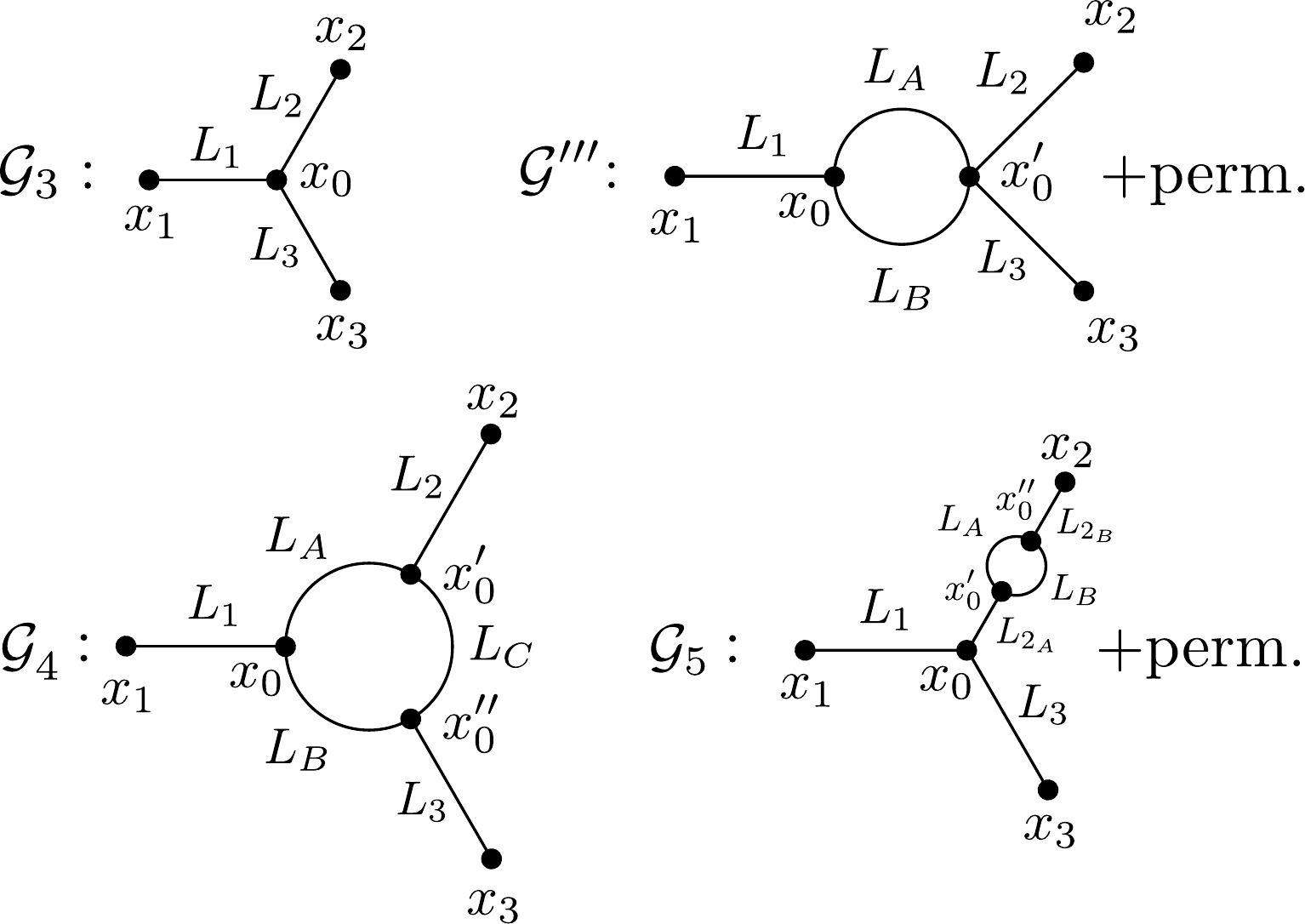}
        \caption{Diagrams considered for the evaluation of the three-point connected correlation function up to one-loop order. Diagram $\mathcal{G}_3$ contributes to $\mathcal{O}(1/M^2)$ while diagrams $\mathcal{G}'''$, $\mathcal{G}_4$ and $\mathcal{G}_5$ contribute to $\mathcal{O}(1/M^3)$.}
        \label{fig:diag_C3}
\end{figure}

With these ingredients, we can compute the observables we are interested in this work, that are the two-point connected and disconnected correlation functions, together with the three-point connected correlation function. In particular, we will denote with 
\begin{equation}
    \mathcal{C}_2(\mathcal{G};\vec{L})\,, \qquad \mathcal{D}_2(\mathcal{G};\vec{L})\qquad   \text{and }\qquad   \mathcal{C}_3(\mathcal{G};\vec{L})\,
\end{equation}
respectively, the two-point connected and disconnected functions and the three-point connected function on generic diagrams $\mathcal{G}$ characterized by a set of line lengths $\vec{L}$. In the BL two sites can only be connected by a unique sequence of adjacent edges, thus the two-point functions are computed on a line and the three-point function on the loop-less topology connecting three sites, respectively represented by $\mathcal{G}_1$ and $\mathcal{G}_3$ depicted in Fig. \ref{fig:diag_CD2} and Fig. \ref{fig:diag_C3}. 
By the definition given in Eq. \eqref{eq:Ptriplet}, the 2-point disconnected correlation function is simply defined as the non-trivial $J=0$ part of the \textit{Ansatz} at null local fields. Thus, from Eq. \eqref{eq:modified_Ansatz_expression}, valid at 0-loop in our expansion:
\begin{equation}\label{eq:disc_bare}
    \mathcal{D}_2(\mathcal{G}_1,L)=-2\,a\,L\,\widetilde{\lambda}^L\,\widetilde{P}_1^2(0)\,.
\end{equation}
In order to compute the two-point connected correlation functions we have to average over the values of the three parameters with the conditions that the two spins belong to the same soft cluster, given by
\begin{equation}
    \prod_{i=1,2}\Theta(|J|-|h_i|)\,\delta(|h_1+h_2\,\sign(J)|)\,,
\end{equation}
thus, all in all, we have
\begin{multline}\label{eq:2point_bare}
    \mathcal{C}_2(\mathcal{G}_1;L) = \!\!\int\! \mathrm{d}J\! \!\int \!\mathrm{d}h_1\!\! \int\! \mathrm{d}h_2\!\! \prod_{i=1,2}\!\widetilde{P}_1(h_i)\Theta(|J|-|h_i|)\times\\\times\delta(|h_1+h_2\sign(J)|)P^{J\neq0}_L(J)
    =\frac{4a}{\widetilde{\rho}}\!\left( \widetilde{P}_1(0) \right)^2\!\widetilde{\lambda}^L+\mathcal{O}\!\left(\!\frac{\widetilde{\lambda}^L}{L}\!\right),
\end{multline}
where $P^{J\neq 0}_L(J) \equiv a L^2 \widetilde{\lambda} L \widetilde{\rho} e^{-\widetilde{\rho}|J|L}$. We remark that the \textit{Ansatz} distribution is valid in the large $L$ limit and, from now on, we neglect higher-order corrections. 

In order to compute the loop-less contribution to the three-point connected function, diagram $\mathcal{G}_3$ in Fig. \ref{fig:diag_C3}, three lines are attached to the internal vertex, yielding an effective Hamiltonian with four effective fields and three effective couplings. As for the two-point function we should average over the distribution of the seven parameters with the condition that the three spins belong to the same cluster, again we refer to \cite{angelini2025criticalexponentsspinglass} for further details. We report only the result obtained by neglecting higher-order terms, since the computation is the same as in Ref.\cite{angelini2025criticalexponentsspinglass}:
\begin{equation}\label{eq:R3bare}
    \mathcal{C}_{3}(\mathcal{G}_3;L_1,L_2,L_3) = \frac{64a^3}{\widetilde{\rho}^3}\left(\widetilde{P}_1(0)\right)^3\widetilde{P}_3(0)\widetilde{\lambda}^{L_1+L_2+L_3}\,.
\end{equation}
In the next sections we will compute corrections to the MF behavior, from which we can estimate the critical exponents.

\section{Main results}\label{sec:Main results}
In this section, we present the main results of the present paper. As in Ref. \cite{angelini2025criticalexponentsspinglass}, we identify a dimensionless ratio between the susceptibilities, whose computation is reported in Sec. \ref{sec:computation of the susceptibilites}, and we look for a fixed point of the corresponding \textit{beta}-function. 

\subsection{Critical exponents}\label{subsec:Critical exponents}
We start here from the expressions of the susceptibilities, computed with the $M$-layer expansion up to one loop order in Sec. \ref{sec:computation of the susceptibilites}, as functions of the \textit{mass}, $m=\xi^{-1}$, related to the distance from the critical point $\tau\equiv-\ln(\widetilde{\lambda}(z-1))$:
\begin{equation}\label{eq:TauFuncOfm}
   \tau\propto m^{\rho-1}\!\!\left(\!1 - \frac{u}{2}\frac{\Gamma\left(\frac{\rho}{\rho-1}\!\right)}{\pi}I_1\left(\left(\frac{m}{\Lambda}\right)^{\rho-1}\right) +\mathcal{O}\left(u^2\right) \!\!\right),
\end{equation}
and then:
\begin{equation}
    \chi_2(m^{\rho-1})\propto \,m^{-(\rho-1)}
    \left(       1  +\mathcal{O}\left(u^2\right)\right)      ,
\label{eq:chi2}
\end{equation}
\begin{multline}
    \chi_3(m^{\rho-1}) \propto m^{-3(\rho-1)} \times\\\!\!
    \times\left(\! 1 -  u\frac{\Gamma\left(\frac{\rho}{\rho-1}\right)}{\pi}I_3\left(\left(\frac{m}{\Lambda}\right)^{\rho-1}\right)  +\mathcal{O}\left(u^2\right)\!\!\right),
\label{eq:chi3}
\end{multline}
\begin{multline}
    \chi_2^\text{dis}(m^{\rho-1}) \propto -m^{-2(\rho-1)}\times \\
    \times\left(       1    -  u\frac{\Gamma\left(\frac{\rho}{\rho-1}\right)}{\pi}I_2\left(\left(\frac{m}{\Lambda}\right)^{\rho-1}\right)  +\mathcal{O}\left(u^2\right) \right)    \,,
\label{eq:chi2dis}
\end{multline}
where $u \equiv g\, m^{5-4\,\rho}$ and $g$ is a $\mathcal{O}(1/M)$ quantity, precisely defined in Sec. \ref{sec:computation of the susceptibilites}, that depends on the microscopic details of the model; $\Lambda$ is the Ultraviolet cutoff introduced due to the presence of a non-zero lattice spacing. Moreover 
\begin{equation}
    I_1(x) =   \int^\infty_{x}\,\mathrm{d}L_a\,\mathrm{d}L_b\frac{L_aL_b\,e^{-(L_a+L_b)}}{(L_a+L_b)^{\frac{1}{\rho-1}+1}}\,,
\end{equation}
\begin{equation}
    I_2(x) = \frac{1}{2}\int^\infty_{x}\,\mathrm{d}L_a\,\mathrm{d}L_b\frac{L_aL_b\,e^{-(L_a+L_b)}}{(L_a+L_b)^{\frac{1}{\rho-1}}}
\end{equation}
and, finally

\begin{equation}
    I_3(x) = \!\!\! \int^\infty_{x} \!\!\!\mathrm{d}L_a\mathrm{d}L_b\mathrm{d}L_c\frac{L_aL_b+L_cL_b+L_aL_c}{(L_a+L_b+L_c)^{\frac{1}{\rho-1}+1}}\,e^{-(L_a+L_b+L_c)}.
\end{equation}
In the following, we want to compute the critical exponents for $\epsilon\equiv\rho-5/4\ll 1$. In the critical limit $m\to0$, $I_2\left(\left(\frac{m}{\Lambda}\right)^{\rho-1}\right)$ and $I_3\left(\left(\frac{m}{\Lambda}\right)^{\rho-1}\right)$ tend to:
\begin{align}
\lim_{m\to0} I_2\!\left(\frac{m^{\rho-1}}{\Lambda^{\rho-1}}\right)
&= \frac{1}{12}\Gamma\!\left(\frac{4\rho-5}{\rho-1}\right),\\
\lim_{m\to0} I_3\!\left(\frac{m^{\rho-1}}{\Lambda^{\rho-1}}\right)
&= \frac{1}{8}\Gamma\!\left(\frac{4\rho-5}{\rho-1}\right).
\end{align}
while $\lim_{m\to0}I_1\left(\left(\frac{m}{\Lambda}\right)^{\rho-1}\right)$ is diverging.
Assuming the scaling laws in Eqs. \eqref{eq:def_K-point_susceptibility} and \eqref{eqeta1}, we can identify the following dimensionless ratio:
\begin{equation}
    \lambda \equiv - m\frac{\chi_2^\text{dis}\chi_3^2}{\chi^4_2}\,,
\end{equation}
not to be confused with the control parameter $\widetilde{\lambda}$. The interesting feature of $\lambda$ is that it can be written as a function of the dimensionless coupling $u$, that is diverging at the critical point, for $\rho\simeq\rho_\text{uc}$. On the other hand, we expect, from the scaling expressions of the susceptibilities, given in Eqs. \eqref{eq:def_K-point_susceptibility} and \eqref{eqeta1}, that $\lambda$ is finite at the critical point. From its critical value, we can estimate critical exponents.
Inserting Eqs. \eqref{eq:chi2},  \eqref{eq:chi3}, \eqref{eq:chi2dis} into the previous equation, up to second order in $u$, we obtain:
\begin{multline}\label{eq:defintion_lambda}
    \lambda \propto u\Bigg(1+\\-u\frac{\Gamma\left(\frac{\rho}{\rho-1}\right)}{\pi}\left(I_2\left(\left(\frac{m}{\Lambda}\right)^{\rho-1}\right)+2I_3\left(\left(\frac{m}{\Lambda}\right)^{\rho-1}\right)\right)\Bigg).
\end{multline}
Following standard field theoretical reasoning \cite{Parisi1988}, to compute the fixed point $\lambda_c$, we appropriately define the following \textit{beta}-function
\begin{equation}
    \beta(\lambda)\equiv m^{\rho-1}\frac{\partial}{\partial m^
    {\rho-1}}\Bigg|_{g\,\text{fixed}}\lambda=\frac{5-4\rho}{\rho-1}u\frac{\partial}{\partial u}\Bigg|_{m^{\rho-1}\,\text{fixed}}\lambda\,,
\end{equation}
and since at the critical point $\lambda$ converges to some $\lambda_c$ we impose $\beta(\lambda_c)=0$; specifically, from Eq. \eqref{eq:defintion_lambda} we get
\begin{equation}\label{eq:lambdac}
    \lambda_c = \frac{3\pi}{\Gamma\left(\frac{\rho}{\rho-1}\right)\Gamma\left(\frac{4\rho-5}{\rho-1}\right)}\,.
\end{equation}
More precisely, we expect that near criticality, where $m$ is small, $\lambda = \lambda_c+ c_1 m^{- \omega}$ with a universal negative exponent $\omega$ that controls the leading finite-size corrections to scaling. This implies:
\begin{equation}\label{eq:correction-to-scaling-def}
\beta(\lambda_c)=0 \ ,\ \ \  \omega=-(\rho-1)\, \beta'(\lambda_c)\,.
\end{equation}
At this point, given the relation between $\tau$ and $m$ given in Eq. \eqref{eq:TauFuncOfm}, we evaluate the exponent $\nu$. We define
\begin{equation}
    z_2(\lambda)\equiv\frac{\partial\tau}{\partial m^{\rho-1}}
\end{equation}
and
\begin{equation}
    c_2(\lambda)\equiv m^{\rho-1}\frac{\partial}{\partial m^{\rho-1}}\Bigg|_{g\,\text{fixed}} \ln z_2(\lambda)\,,
\end{equation}
such that
\begin{equation}\label{eq:definition_nu}
    c_2(\lambda_c)=\frac{1}{\nu(\rho-1)}-1\,.
\end{equation}
Realizing that, at one-loop order, Eq. \eqref{eq:defintion_lambda} reduces to $u\simeq-\lambda$, we can express Eq. \eqref{eq:TauFuncOfm} as a function of $\lambda$ and, given the value $\lambda_c$, we evaluate $c_2(\lambda_c)$
\begin{equation}\label{eq:c2}
    c_2(\lambda_c)=\frac{5-4\rho}{4(\rho-1)}\,.
\end{equation}
Notice that even if for $\epsilon=\rho-5/4\ll1$ the integral $I_1\left(\left(\frac{m}{\Lambda}\right)^{\rho-1}\right)$ is diverging in the $m\to0$ limit, taking the derivative, as prescribed by the definition of $z_2(\lambda)$, the result is no more diverging and can be expressed as a function of $\rho$. Using Eqs. \eqref{eq:definition_nu} and \eqref{eq:c2} we obtain the expression for $\nu$ to first order in $\epsilon$
\begin{equation}
    \nu=4+\mathcal{O}(\epsilon^2)\,.
\end{equation}
A similar expression can be found to first order in $\epsilon$ for the correction-to-scaling critical exponent. From Eq. \eqref{eq:correction-to-scaling-def}:
\begin{equation}
    \omega=-4\epsilon+\mathcal{O}(\epsilon^2)\,.
\end{equation}
We recall, from Ref. \cite{angelini2025criticalexponentsspinglass}, that the correction-to-scaling exponent $\omega$ for the SR $D$-dimensional model is $\omega_\text{sr}=D-8$ at one-loop order, thus, given the relation between $\rho$ and $D$ in Eq. \eqref{eq:Dim_SR_LR_correspondence}, one can easily verify that $\omega_\text{sr}$ and $\omega$ satisfy the correspondence in Eq. \eqref{eq:correspondence_SR-LR} only above the upper critical dimension $D_\text{uc}=8$, where $\eta_\text{sr}=0$. 
On the contrary, the same correspondence at one-loop order is verified for the exponent $\nu$ above and below $D_\text{uc}=8$.

We leave the analysis of the anomalous dimensions $\eta$ and $\overline{\eta}$ to the next section.

\subsection{Anomalous dimensions for the long-range model}\label{subsec:Anomalous dimensions for the long range model}
The last two critical exponents to compute are the anomalous dimensions, $\eta$ and $\overline{\eta}$, for which we dedicated this Section. Indeed, a non-trivial result from the $M$-layer construction, applied to the LR version of the spin glass model we are interested in, is the $\epsilon$-expansion for these two exponents, which reveals that $\eta$  sticks to the MF value $\eta=0$. This result is consistent with field-theoretical predictions, as the Gaussian interaction in these LR-type models is non-local \cite{Fisher72Critical,paulos2016conformal}. As we will see in the following, in the $M$-layer construction, the two point correlation in the momentum space at small values of $k$ can be written as:
\begin{equation}
    \widehat{C}_2(k)=\left(\widehat{C}_2(0)^{-1}+ \, \gamma \,k^{\rho-1}\right)^{-1}\,.
\label{generic}
\end{equation}
In the above expression  $\widehat{C}_2(0)^{-1}$ has diverging $1/M$ corrections while the coefficient $\gamma$ of $k^{\rho-1}$ does not depend on $M$. Indeed, the $1/M$ expansion does not generate terms proportional to $k^{\rho-1}$, only terms proportional to $k^2$ and higher orders.
It follows that according to the definition of $\xi$, written in Eq. \eqref{eq:def_corr_length}, we have
\begin{equation}
   m^{1-\rho}\equiv \xi^{\rho-1} =   \widehat{C}_2(0) \ .
\end{equation}
Given that $\chi_2= \widehat{C}_2(0)$ we have therefore 
\begin{equation}
    \chi_2 \propto \xi^{\rho-1}=m^{1-\rho}\,,
\end{equation}
which implies that Eq. (\ref{eq:chi2}) has vanishing corrections at all orders in $u$.
Comparison with Eq. (\ref{eq:X2_scal}) implies the remarkable property:
\begin{equation}
   \eta=0 \ \ \quad \forall  \, \rho\in (1,3) \,. 
\end{equation}
Furthermore, the expression of Eq. (\ref{generic}) implies that in the critical region $r=\mathcal{O}(\xi)$ the correlation is described by the {\it same} scaling form that holds in MF:
\begin{equation}
     \widehat{C}_2(k)=\frac{1}{m^{\rho-1}+ \gamma \, k^{\rho-1}}\,,
\end{equation}
which implies that also for $\rho>\rho_\text{uc}$ the correlation at the critical point decays at large $r$ as 
\begin{equation}
    C_2(0,r)\sim r^{\rho-2}
\end{equation}
again in agreement with $\eta=0$. Close to the critical point, the above non-integrable power-law is cut off on the scale $r=\mathcal{O}(\xi)$ by the integrable power-law $ C_2(0,r)\sim r^{-\rho} $.

In order to compute the exponent $\overline{\eta}$, we consider the scaling law $\chi_2^\text{dis} \propto -\xi^{2(\rho-1)-\overline{\eta}} $ and we define
\begin{equation}
    Q_2^\text{dis}(\lambda_c)\equiv\frac{\partial\ln\left(-\chi_2^\text{dis}(\lambda)\right)}{\partial m^{\rho-1}}\Bigg|_{\lambda_c}=\frac{\overline{\eta}}{\rho-1}-2\,.
\end{equation}
Given the expression of $\chi_2^\text{dis}(m)$ in Eq. \eqref{eq:chi2dis} and $\lambda_c$ we compute the first order correction to $\overline{\eta}$ 
\begin{equation}
    \overline{\eta}=\rho-\frac{5}{4}\,,
\end{equation}
from which, using $\epsilon=\rho-5/4$
\begin{equation}
    \overline{\eta}=\epsilon+\mathcal{O}(\epsilon^2)\,.
\end{equation}
The result is that while $\eta=0$, for all $\rho$, $\overline{\eta}\neq 0$ for $\rho > \rho_\text{uc}$.

Notice also that the approximate correspondence between the finite-dimensional and the corresponding LR model, described in Sec. \ref{subsec:Scaling laws}, is exact at one-loop order for $\nu$, $\eta$, and $\overline{\eta}$. Indeed, expressing $\epsilon\equiv\rho-5/4$ in terms of $\epsilon_\text{sr}\equiv8-D$, using Eq. \eqref{eq:Dim_SR_LR_correspondence}, it is possible to show that
\begin{equation}
	D \nu_\text{sr} = \nu \,,
\end{equation}
\begin{equation}
	 \frac{D + 2 - \eta_\text{sr}}{D} = \rho - \eta \,,
\end{equation}
and
\begin{equation}
	\frac{\eta_\text{sr} - \overline{\eta}_\text{sr} +2}{D} = \eta  - \overline{\eta} + \rho -1 \,.
\end{equation}
The same correspondence holds at one-loop order for the Edwards-Anderson model without an external field at finite temperature \cite{harris1976critical,Kotliar_1983,Martin-Mayor_2012}.

\section{The method: $M$-layer construction}\label{sec:The method: M-layer construction}

\subsection{Description of the Method}\label{sec:Description of the Method}

The $M$-layer construction applied to a finite-dimensional lattice, as presented in \cite{Altieri_2017,Angelini_2024Ising,angelini2025bethe}, is generalized here when the \textit{original lattice} is the one-dimensional LR lattice, defined in Sec. \ref{sec: the model}. 

Once the original lattice is generated with the random procedure defined in Sec. \ref{sec: the model}, $M-1$ additional identical copies (or \textit{layers}) of it are created. The links in the $M$ layers are then reshuffled with the following random rewiring procedure. Each copy is identified by an index, $\alpha = 1, 2, \ldots, M$, starting with the one corresponding to the original lattice, $\alpha = 1$. Each node of this $M$-layered lattice will thus be identified by two indices: For instance, $i_\alpha$ refers to node $i$ in layer $\alpha$. Suppose now that $i_\alpha$ is connected to the node $j_\alpha$ in the same layer $\alpha$. It is thus possible to identify two sets of nodes, $\{i_\alpha\}_{\alpha=1,\ldots,M}$ and $\{j_\alpha\}_{\alpha=1,\ldots,M}$, connected by $M$ edges $(i_\alpha, j_\alpha)$. The random rewiring procedure generates inter-layer connections by permuting the links between these sets, giving rise to a new set of edges:
\begin{equation}
    (i_\alpha, j_{\pi(\alpha)})
\end{equation}
for $\alpha = 1, \ldots, M$, where $\pi(\alpha)$ is one of the possible $M!$ permutations of the set $(1, 2, \ldots, M)$. Notice that, with this construction, the connectivity of each node remains $z$. Repeating this random rewiring procedure for all the remaining edges, an instance of the so-called \textit{M-layer lattice} is obtained. 

As in the finite-dimensional construction \cite{Angelini_2024Ising}, it is possible to understand some properties of the resulting layered graph. First, the original lattice is recovered for $M=1$. Also, in the limit $M\to\infty$ the probability to close a topological loop goes to zero. This can be intuitively understood by noting that, as the number of layers increases, the probability of connecting one node to another specific node on a generic layer is exactly $1/M$ after the rewiring. In this sense, the parameter $1/M$ can be varied to tune the probability of having topological loops in the system, which affects the critical behavior. 

The path between any two sites can be projected onto original one-dimensional lattice, yielding the NBP, as described above. Specifically, in this context, the fact that a projected path is \quotationmarks{non-backtracking} is reflected in the fact that given a step from $x_1$ to $x_2$, the adjacent step cannot be from $x_2$ to $x_1$.

Next, we will show how perturbative expansion, in the small parameter $1/M$, can be computed for interesting correlation functions, necessary to compute the critical exponents of the original one-dimensional LR lattice.

\subsubsection{Random walks on long-range graphs}\label{sec:random_walks}

Unfortunately, the computation of the number of NBP for the fixed connectivity $M$-layer graph, described in Sec. \ref{sec:Description of the Method} is not easily done. However, since we are interested in the large length asymptotic expressions, we can make an approximate argument to obtain it for the ensemble of LRRGs with fixed connectivity $z$, considered in this paper, defined in Sec. \ref{sec: the model}.

Consider the number of NBP with $L$ steps between two nodes at $x_1$ and $x_2$ on a given realization of the graph $G$, $\mathcal{N}^{(G)}_L(x_1,x_2)$. It obeys the following equation:
\begin{equation}
    \mathcal{N}_L^{(G)}(x_1,x_2)=\sum_{x_0\in\mathbb{Z}^1}\mathcal{N}^{(G)}_{L-1}(x_1,x_0)\,\boldsymbol{1}_{\mathcal{E}(G)}[(0,2)]\,,
\end{equation}
where $\boldsymbol{1}_{\mathcal{E}(G)}[(i,j)]$ is the indicator function, equal to one if the edge $(i,j)$ is present in graph $G$ and zero otherwise. The quantity we are interested in is the average over $G_z$, that is $\mathcal{N}_L(x_1,x_2)$. In general, when averaging, it is not correct to factorize the two terms in the RHS, since the number of NBP of length $L-1$ arriving at position $x_0$ is correlated with the probability of occurrence of the edges adjacent to $x_0$. However, we expect that, in the large distances regime, the factorization is approximately correct and we obtain:
\begin{equation}\label{eq:recursive_NBP}
    \mathcal{N}_L(x_i,x_j)=\sum_{x_w\in\mathbb{Z}^1}\mathcal{N}_{L-1}(x_i,x_w)\,P(x_w-x_j)\,\ ,
\end{equation}
where $P(x_i-x_j)$, defined in Sec. \ref{sec: the model}, is the probability that the edge between two nodes at positions $x_i$ and $x_j$ is present in $G$. As shown in Fig. \ref{fig:random_walk}, in the large distances regime $P(x_i-x_j)=\mathcal{A}|x_i-x_j|^{-\rho}$ where $1<\rho<3$ and $\mathcal{A}$ is a normalization factor. To solve the self-consistency equation for $\mathcal{N}_L(x_i,x_j)$, Eq. \eqref{eq:recursive_NBP}, we employ the Fourier transform of the convolution, using the following convention:
\begin{equation}
    \widehat{f}(k)=\sum_{x\in\mathbb{Z}^1}f(x)e^{ikx},\quad f(x)=\int_{\left[-\pi,\pi\right]}\frac{\mathrm{d}k}{2\pi}\widehat{f}(k)e^{-ikx},
\end{equation}
that implies the expression of the Dirac delta function in the reciprocal space:
\begin{equation}
    2\pi \,\delta(k)=\sum_{x\in\, \mathbb{Z}}e^{ikx}\,.
\end{equation}
The Fourier transform of Eq. \eqref{eq:recursive_NBP} reads
\begin{equation}\label{eq:recursive_NBP_Fourier}
    \widehat{\mathcal{N}}_L(k)=\widehat{\mathcal{N}}_{L-1}(k)\,\widehat{P}(k)\,,
\end{equation}
where we defined $\widehat{N}_L(k)$ from :
\begin{equation}
    \widehat{\mathcal{N}}_L(k_i,k_j)=2\pi\,\delta(k_i+k_j)\widehat{\mathcal{N}}_L(k_i)\,,
\end{equation}
and the same for $\widehat{P}(k)$. Now, since we are interested in large $L$ asymptotic expressions we consider $k\approx0$ and, using the expression of $P(r)$, we obtain
\begin{equation}
    \widehat{P}(k)-\widehat{P}(0)=-2\mathcal{A}\,k^{\rho-1}\sum_{r=1}^{\infty}\frac{1}{r^\rho}\left(1-\cos(r)\right)\equiv-\alpha_\rho \,k^{\rho-1}\,,
\end{equation}
with $\alpha_\rho$ a $\rho$-dependent quantity, well-defined for $\rho>1$.
Plugging the asymptotic expression for $\widehat{P}(k)$ in Eq. \eqref{eq:recursive_NBP_Fourier} we can solve for $\widehat{\mathcal{N}}_L(k)$
\begin{equation}
    \widehat{\mathcal{N}}_L(k)=z(z-1)^{L-1}\,\exp(-\alpha_\rho\,k^{\rho-1}L)\,,
\end{equation}
which is the Fourier transform of the number of NBP of length $L$, with momentum $k$, while the factor $z(z-1)^{L-1}$ is a normalization given by the zero momentum (large distance) limit, where the number of NBP is simply $z(z-1)^{L-1}$. We remark that we set the lattice spacing $a_l=1$. In general, a factor $a_l$ would appear in front of the exponential and a factor $a_l^{\rho-1}$ at the exponent.

\begin{figure}[h]
    \centering

    \includegraphics[width=1.\linewidth]{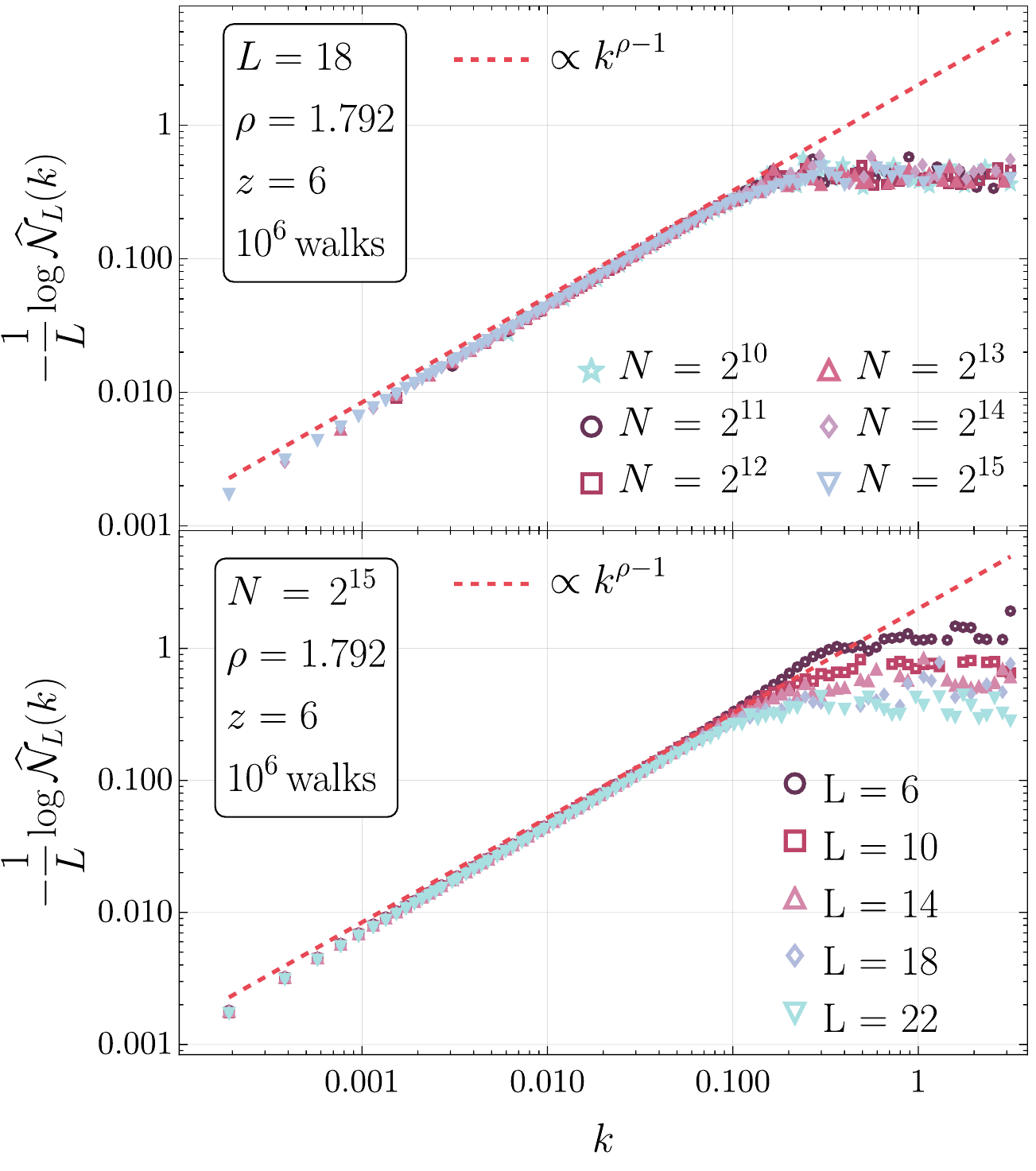}

    \caption{ Numerical evaluations of the number of NBP on one-dimensional LRRGs with fixed connectivity $z=6$ and decaying exponent $\rho=1.792$. On both figures $10^6$ NBP, with a fixed length $L$, are performed on the 1000 LRRGs previously generated with the Monte Carlo procedure of Ref. \cite{Martin-Mayor_2012}. Top: Fourier transform of the number of NBP with fixed length $L=18$ for different system sizes $N$, compared with the expected curve $k^{\rho-1}$. 
    Bottom: Fourier transform of the number of NBP with fixed system size $N=16384$ for different lengths $L$, compared with the expected curve $k^{\rho-1}$. Notice that both curves could be obtained for different values of the connectivity $z$ and the decaying exponent $\rho$. While the same behavior is expected, decreasing $\rho$ towards $\rho=1$ increases both the equilibration time and the finite-size effects. }
    \label{fig:random_walk2}
\end{figure}

\subsection{Computation of the susceptibilities}\label{sec:computation of the susceptibilites}

In this Section, we present the technical steps, prescribed by the $M$-layer construction, to compute the susceptibilities reported in Sec. \ref{sec:Main results}. 

To avoid all the null inputs of the three observables ($\mathcal{P}_2(x_1,x_2;0)$, $P^\text{dis}(x_1,x_2;0,0)$ and $\mathcal{P}_3(x_1,x_2,x_3;0)$) we define the two-point connected and disconnected functions together with the three-point connected functions respectively as
\begin{align}
\mathcal{P}_2(x_1,x_2;0)
&\equiv C_2(x_1,x_2), \\
P^{\text{dis}}(x_1,x_2;0,0)
&\equiv D_2(x_1,x_2), \\
\mathcal{P}_3(x_1,x_2,x_3;0)
&\equiv C_3(x_1,x_2,x_3),
\end{align}
averaged over the disorder and the rewirings on the $M$-layer lattice. 

As prescribed by the $M$-layer construction \cite{Altieri_2017,Angelini_2024Ising} for each observable we need to: i) identify relevant diagrams; ii) compute the weight $W$ (as a function of $M$), $\mathcal{N}$ and the associated symmetry $S$ factors; iii) compute the observable on the identified diagrams and iv) sum the contributions. The third step is the only model-dependent one, while the computation of the number of NBP is strictly dependent on the graph used in the lattice model. Let us start with the connected and disconnected two-point functions, $C_2(x_1,x_2)$ and $D_2(x_1,x_2)$, for which the same topological diagrams are considered. 
\vspace{0.2cm}

\textit{Observables: $C_2(x_1,x_2)$ and $D_2(x_1,x_2)$}

\begin{itemize}
    \item[\ding{182}] Identification of relevant diagrams

The first diagram to consider is the bare line, denoted as $\mathcal{G}_1$. In the presence of a loop, we analyze the diagram comprising four lines and two vertices of degree three, with the two internal lines forming a closed loop. We denote this diagram as $\mathcal{G}_2$ (refer to Fig. \ref{fig:diag_CD2}).  

The reasons for including only these two diagrams can be found in Ref. \cite{angelini2025criticalexponentsspinglass}: the determination of the diagrams that give the dominant contribution to a chosen observable is independent on the original lattice and in this particular case, the dominant diagrams are the same for the LR lattice as for the finite-dimensional regular lattice analyzed in Ref. \cite{angelini2025criticalexponentsspinglass}.
%they are independent of the specific problem considered. 
Thus, we will only focus on $\mathcal{G}_1$ and $\mathcal{G}_2$ for the two-point functions.  
    
    \item[\ding{183}] $W$, $\mathcal{N}$ and $S$ factors 
    
  Specifically, given a generic diagram $\mathcal{G}$ with lengths $\Vec{L}$ and external vertices at $x_1,...,x_q$, we need to compute $\mathcal{N}(\mathcal{G};\Vec{L};x_1,...,x_q)$, that we define as the number of realizations of the topological diagram $\mathcal{G}$ with set of lengths $\Vec{L}$ and $q$ external points $x_1,...,x_q$ on the original LR model. This number is related to the number of NBP connecting the starting and ending points of each line. 
  
  As in the finite-dimensional regular lattice case, the same symmetry factors $S(\mathcal{G})$ are associated with each diagram.
  The factors $W$, $\mathcal{N}$ and $S$ result to be the following for the two-point diagrams:
  
    	Diagram $\mathcal{G}_1$\begin{itemize}
		\item[$\bullet$] $W(\mathcal{G}_1) = \frac{1}{M}$;
		\item[$\bullet$] $\mathcal{N}(\mathcal{G}_1;L;x_1,x_2) = \mathcal{N}_L(x_1,x_2)$;
		\item[$\bullet$] $S(\mathcal{G}_1) = 1$.
	\end{itemize}
	
	Diagram $\mathcal{G}_2$\begin{itemize}
		\item[$\bullet$] $W(\mathcal{G}_2) = \frac{1}{M^2}$;
		\item[$\bullet$] $\mathcal{N}(\mathcal{G}_2;\Vec{L};x_1,x_2) = z^2\left(\frac{z!}{(z-3)!}\right)^2 \times \\ 
        \times \!\! \sum\limits_{x_0,x_0'}\!\! \frac{\mathcal{N}_{L_1}(x_1,x_0)}{z^2} \frac{\mathcal{N}_{L_A}(x_0,x_0')}{z^2}\frac{\mathcal{N}_{L_B}(x_0,x_0')}{z^2} \frac{\mathcal{N}_{L_2}(x_0',x_2)}{z^2}$;
		\item[$\bullet$] $S(\mathcal{G}_2) = 2$\,,
	\end{itemize}
		where $\Vec{L} = (L_1, L_A, L_B, L_2)$.

    \item[\ding{184}] Compute the observable on the identified diagram
    
    We introduce a generalized notation for the observables on the BL, using the superscript \quotationmarks{lc}: \begin{equation} 
    \mathcal{C}_2^\text{lc}(\mathcal{G};{\Vec{L}}) \qquad \quad \text{and} \qquad \quad \mathcal{D}_2^\text{lc}(\mathcal{G};{\Vec{L}})\,,
\end{equation} 
which stands for \quotationmarks{line-connected}, a definition that is used in the the $M$-layer framework to avoid double counting of loop contributions (see Ref. \cite{Altieri_2017} for more details). 
We again refer to \cite{angelini2025criticalexponentsspinglass} for the computation of the observables on the given diagrams, which are completely identical, since the model is the same. We already reported the loop-less results in Eqs. \eqref{eq:2point_bare}, \eqref{eq:disc_bare} and \eqref{eq:R3bare}, we add the contributions for the loop-corrections:
\begin{multline}
        \mathcal{C}_{2}^\text{lc}(\mathcal{G}_2;\Vec{L})=-\frac{128a^4}{\widetilde{\rho}^3}\left(\widetilde{P}_1(0)\right)^2\left(\widetilde{P}_3(0)\right)^2\times\\
        \times\frac{L_AL_B}{L_A+L_B}\widetilde{\lambda}^{L_1+L_A+L_B+L_2}\,,\label{eq:R2loop}
\end{multline}
\begin{multline}\label{eq:D2loop}
    \mathcal{D}_{2}^\text{lc}(\mathcal{G}_2;\Vec{L})=64 \frac{a^4 \left(\widetilde{P}_3(0)\right)^2\,\left(\widetilde{P}_1(0)\right)^2}{{\widetilde{\rho}^2}}\times \\
    \times\frac{{\widetilde{\lambda}^{L_1 + L_2 + L_A+ L_B}}}{{L_A+ L_B}} L_AL_B(L_1 + L_2 + L_A+ L_B)\,,
\end{multline}

    \item[\ding{185}] Sum of the contributions
    
    The perturbative expression of the two-point connected function on the $M$-layer lattice, averaged over the rewirings, $C_2(x_1,x_2)$, up to one-loop contributions, is the sum over the two relevant diagrams:
\begin{multline}
    C_2(x_1,x_2)\,=\frac{1}{M}\sum_L\mathcal{N}_L(x_1,x_2)\mathcal{C}_{2}^\text{lc}(\mathcal{G}_1;L) \\+\frac{1}{2M^2}\sum_{\Vec{L}}\mathcal{N}(\mathcal{G}_2;\Vec{L};x_1,x_2)\mathcal{C}_{2}^\text{lc}(\mathcal{G}_2;\Vec{L})+\mathcal{O}\left(\frac{1}{M^3}\right),\label{eq:2point}
\end{multline}
where $\Vec{L}=(L_1,L_A,L_B,L_2)$. Analogously, we can write the same expansion for the disconnected function
\begin{multline}
    D_2(x_1,x_2)\,=\frac{1}{M}\sum_L\mathcal{N}_L(x_1,x_2)\mathcal{D}_{2}^\text{lc}(\mathcal{G}_1;L)\\+\frac{1}{2M^2}\sum_{\Vec{L}}\mathcal{N}(\mathcal{G}_2;\Vec{L};x_1,x_2)\mathcal{D}_{2}^\text{lc}(\mathcal{G}_2;\Vec{L})+\mathcal{O}\left(\frac{1}{M^3}\right)\,,\label{eq:2pointdisconnected}
\end{multline}
where we remark that the only difference is the observable computed on a given diagram.
\end{itemize}
The same notation is applied to the three-point function: $\mathcal{C}_3^\text{lc}(\mathcal{G};{\Vec{L}})$. Similar arguments are valid for it and we only report here the final result together with the values of the observables on the given diagrams:
\begin{widetext}
\begin{multline}
    C_3(x_1,x_2,x_3)=\frac{1}{M^2}\!\!\sum_{L_1,L_2,L_3}\!\!\!\!\mathcal{N}(\mathcal{G}_3;L_1,L_2,L_3;x_1,x_2,x_3)\mathcal{C}_{3}^\text{lc}(\mathcal{G}_3;\{L_1,L_2,L_3\})\\
    +\frac{1}{M^3}\sum_{\Vec{L}'}\mathcal{N}(\mathcal{G}_4;\Vec{L}';x_1,x_2,x_3)\mathcal{C}_{3}^\text{lc}(\mathcal{G}_4;\Vec{L}')
    + \frac{1}{2M^3}\sum_{\Vec{L}''}\mathcal{N}(\mathcal{G}_5;\Vec{L}'';x_1,x_2,x_3)\mathcal{C}_{3}^\text{lc}(\mathcal{G}_5;\Vec{L}'') +\text{perm.}+ \mathcal{O}\left(\frac{1}{M^4}\right),\label{eq:3point}
\end{multline}
with $\Vec{L}'=(L_1,L_2,L_3,L_A,L_B,L_C)$, $\Vec{L}''=(L_1,L_{2_A},L_A,L_B,L_{2_B},L_3)$, the considered diagrams are shown in Fig. \ref{fig:diag_C3},
\begin{equation}\label{eq:R3loopOnExtLeg}
        \mathcal{C}_{3}^\text{lc}(\mathcal{G}_5;\Vec{L''})=-\frac{2048a^6}{\widetilde{\rho}^5}\left(\widetilde{P}_1(0)\right)^3\left(\widetilde{P}_3(0)\right)^3\frac{L_AL_B}{L_A+L_B}\widetilde{\lambda}^{L_1+L_{2_A}+L_{2_B}+L_A+L_B+L_3}\,,
\end{equation}
\begin{equation}\label{eq:R3loop}
        \mathcal{C}_{3}^\text{lc}(\mathcal{G}_4;\Vec{L'})=-\frac{2048a^6}{\widetilde{\rho}^5}\left(\widetilde{P}_1(0)\right)^3\left(\widetilde{P}_3(0)\right)^3\frac{L_AL_B+L_CL_B+L_AL_C}{L_A+L_B+Lc}\widetilde{\lambda}^{L_1+L_2+L_3+L_A+L_B+L_C}\,.
\end{equation}
\end{widetext}
In order to compute the corresponding susceptibilities we first evaluate the complete expressions of the correlation functions given in Eqs. \eqref{eq:2point}, \eqref{eq:2pointdisconnected} and \eqref{eq:3point} using the results for the observables computed on given diagrams, $\mathcal{C}_2^{lc}$, $\mathcal{D}_2^{lc}$ and $\mathcal{C}_3^{lc}$. We define the Fourier transforms of such expressions as

\begin{multline}
    \widehat{C}_2(k_1,k_2)\equiv a_l^2\sum_{x_1,x_2}e^{ik_1\,x_1+ik_2\,x_2}C_2(x_1,x_2)\\\equiv 2\pi\,\delta(k_1+k_2)\widehat{C}_2(k_1)\,,
\end{multline}
\begin{multline}
    \widehat{D}_2(k_1,k_2)\equiv a_l^{2}\sum_{x_1,x_2}e^{ik_1\,x_1+ik_2\,x_2}D_2(x_1,x_2)\\\equiv 2\pi\,\delta(k_1+k_2)\widehat{D}_2(k_1)\,,
\end{multline}
\begin{multline}
    \widehat{C}_3(k_1,k_2,k_3)\equiv a_l^3\!\!\sum_{x_1,x_2,x_3}\!\!\!\!e^{ik_1\,x_1+ik_2\,x_2+ik_3\,x_3}C_3(x_1,x_2,x_3)\\\equiv 2\pi\,\delta(k_1+k_2+k_3)\widehat{C}_3(k_1,k_2)\,,
\end{multline}
where we made the lattice spacing $a_l$ explicit. The corresponding expressions are
\begin{widetext}
\begin{multline}\label{eq:completeC2}
    \widehat{C}_2(k)=4\,\widetilde{\rho}\,a_l\,\frac{C\,B^2}{A}\left(\sum_{L=1}^\infty e^{-\left(k^{\rho-1}+\tau\right)L}\right)\Bigg( 1+\\
    -16\,A\sum_{L=1}^\infty e^{-\left(k^{\rho-1}+\tau\right)L}\sum_{L_A,\,L_B}\frac{L_AL_B}{L_A+L_B}\int\frac{\mathrm{d}q}{2\pi} e^{-\left(q^{\rho-1}+\tau\right)L_A-\left((q-k)^{\rho-1}+\tau\right)L_B}\Bigg)\,,
\end{multline}

\begin{multline}\label{eq:completeD2}
    \widehat{D}_2(k)=-2\,\widetilde{\rho}^2\,a_l\,\frac{C\,B^2}{A}\left(\sum_{L=1}^\infty L\,e^{-\left(k^{\rho-1}+\tau\right)L}\right)\Bigg( 1+\\
    -32\,A\left(\sum_{L=1}^\infty e^{-\left(k^{\rho-1}+\tau\right)L}\right)\sum_{L_A,\,L_B}\frac{L_AL_B}{L_A+L_B}\int\frac{\mathrm{d}q}{2\pi} e^{-\left(q^{\rho-1}+\tau\right)L_A-\left((q-k)^{\rho-1}+\tau\right)L_B}+\\
    -16\,A\frac{\left(\sum_{L=1}^\infty e^{-\left(k^{\rho-1}+\tau\right)L}\right)^2}{\sum_{L=1}^\infty L\,e^{-\left(k^{\rho-1}+\tau\right)L}}\sum_{L_A,\,L_B}L_A\,L_B\int\frac{\mathrm{d}q}{2\pi} e^{-\left(q^{\rho-1}+\tau\right)L_A-\left((q-k)^{\rho-1}+\tau\right)L_B}\Bigg)\,,
\end{multline}

\begin{multline}\label{eq:completeC3}
    \widehat{C}_3(k_1,k_2)=64\,\widetilde{\rho}\,
    a_l \frac{C\,B^3}{A}\prod_{i=1,2}\left(\sum_{L_i=1}^\infty \,e^{-\left(k_i^{\rho-1}+\tau\right)L_i}\right)\left(\sum_{L_3=1}^\infty \,e^{-\left((k_1+k_2)^{\rho-1}+\tau\right)L_3}\right)\Bigg( 1+\\
    -16\,A\sum_{L_A,\,L_B,\,L_C}\frac{L_A\,L_B+L_A\,L_C+L_B\,L_C}{L_A+L_B+L_C}\int\frac{\mathrm{d}q}{2\pi} e^{-\left(q^{\rho-1}+\tau\right)L_A-\left((q-k_1)^{\rho-1}+\tau\right)L_B-\left((q+k_2)^{\rho-1}+\tau\right)L_C}\\
    -16\,A\sum_{i=1,2,3}\left(\sum_{L=1}^\infty e^{-\left(k_i^{\rho-1}+\tau\right)L}\right)\sum_{L_A,\,L_B}\frac{L_AL_B}{L_A+L_B}\int\frac{\mathrm{d}q}{2\pi} e^{-\left(q^{\rho-1}+\tau\right)L_A-\left((q-k_i)^{\rho-1}+\tau\right)L_B}\Bigg)\,,
\end{multline}
\end{widetext}
where we defined the following quantities
\begin{multline}
    A\equiv \frac{1}{M}\left(\frac{z!}{(z-3)!}\right)^2\frac{\left(\widetilde{P}_3(0)\right)^2}{\widetilde{\rho}^2}\,\left(\alpha_\rho\right)^{\frac{1}{1-\rho}}\,a^3, \\
    B\equiv\frac{1}{M}\frac{z!}{(z-3)!}\frac{\widetilde{P}_3(0)\,\widetilde{P}_1(0)}{\widetilde{\rho}^2}\,a^2, \quad     C\equiv\left(\alpha_\rho\right)^{\frac{1}{1-\rho}}\,,
\end{multline}

At this point, we can compute the susceptibilities from 

\begin{equation}\label{eq:def_chi2_tau}
    \sum_{x}C_2(x,0)=\frac{1}{a_l}\widehat{C}_{2}(0) = \chi_2\,,
\end{equation}

\begin{equation}\label{eq:def_chi2dis_tau}
    \sum_{x,y}C_3(x,y,0)=\frac{1}{a_l^2}\,\widehat{C}_{3}(0,0) = \chi_3\,,
\end{equation}

\begin{equation}\label{eq:def_chi3_tau}
    \sum_{x}D_2(x,0)=\frac{1}{a_l}\widehat{D}_{2}(0) = \chi^\text{dis}_2\,,
\end{equation}
and the final result is
\begin{equation}
    \chi_2(\tau) = 4\frac{\widetilde{\rho}}{\tau}C\,\frac{B^2}{A}\left(       1    -    16\,A\,\tau^{\frac{1}{\rho-1}-4}\frac{\Gamma\left(\frac{\rho}{\rho-1}\right)}{\pi}\,I_1\left(\frac{\tau}{\Lambda^2}\right)         \right)\,,
\end{equation}
\begin{multline}
   \chi_3(\tau) = 64 \frac{\widetilde{\rho}}{\tau^3}C\,\frac{B^3}{A}\Bigg(  1-32\,A\,\tau^{\frac{1}{\rho-1}-4}\frac{\Gamma\left(\frac{\rho}{\rho-1}\right)}{\pi}\,I_3\left(\frac{\tau}{\Lambda^2}\right)\\-48\,A\,\tau^{\frac{1}{\rho-1}-4}\frac{\Gamma\left(\frac{\rho}{\rho-1}\right)}{\pi}\,I_1\left(\frac{\tau}{\Lambda^2}\right)\Bigg)\,,
\end{multline}
\begin{multline}
    \chi_2^\text{dis}(\tau)= -2\frac{\widetilde{\rho}^2}{\tau^2}C\,\frac{B^2}{A}\Bigg( 1+\\-32\,A\,\tau^{\frac{1}{\rho-1}-4}\frac{\Gamma\left(\frac{\rho}{\rho-1}\right)}{\pi}\,\left(  
 I_1\left(\frac{\tau}{\Lambda^2}\right) + I_2\left(\frac{\tau}{\Lambda^2}\right)  \!\right)\!\!\Bigg), 
\end{multline}
where the integrals $I_1(x),\,I_2(x)$ and $I_3(x)$ are defined in Sec. \ref{subsec:Critical exponents}.

To regularize the divergence of $I_1$ we should rewrite the susceptibilities in terms of the mass, $m=\xi^{-1}$. To this aim we use the definition of the correlation length in Eq. \eqref{eq:def_corr_length} and we write here $\widehat{C}_2^{-1}(k)$
\begin{multline}
    \widehat{C}_2^{-1}(k)\propto k^{\rho-1}+\tau+\\+16A\!\!\int\!\! \frac{\mathrm{d}q}{2\pi}\!\!\int \!\mathrm{d}L_a\mathrm{d}L_b\frac{L_aL_be^{-q^{\rho-1}L_a-(q+k)^{\rho-1}L_b-\tau(L_a+L_b)}}{L_a+L_b}\,,
    \label{C20}
\end{multline}
neglecting higher order corrections in $A\propto 1/M$. 
The last term depends on $k$, for $k=0$ it yields the $\mathcal{O}(1/M)$ correction to $\widehat{C}_2^{-1}(0)$.
The second term can be expanded in powers of $k$; the first non-zero term in the expansion is proportional to $k^2$, which,  at small values of $k$, is negligible  with respect to $k^{\rho-1}$ as soon as $\rho<3$. In other words, the coefficient of the $k^{\rho-1}$ sticks to one and it is not altered by the $1/M$ expansion. As we explained in Sec. \ref{subsec:Anomalous dimensions for the long range model}, this implies that 
\begin{equation}
    m^{\rho-1}=\widehat{C}_2^{-1}(0)=\chi_2^{-1} \ .
\end{equation}
Expression (\ref{C20}) gives $\widehat{C}_2^{-1}(0)$ in powers of $\tau$, plugging it into the above equation and inverting it, we obtain $\tau$ as a function of $m$:
\begin{equation}\label{eq:tau_m}
    \tau \propto m^{\rho-1}\left(1-16\,A\,m^{5-4\rho}\frac{\Gamma\left(\frac{\rho}{\rho-1}\right)}{\pi}I_1(m^{\rho-1}/\Lambda^2)\right)\,.
\end{equation}
We can insert Eq. \eqref{eq:tau_m} inside the expressions of the susceptibilities, obtaining them as a function of $m$ instead of $\tau$
\begin{multline}
    \chi_3(m^{\rho-1}) \propto \\m^{-3(\rho-1)} 
    \left(       1    -  32A\,m^{5-4\rho}\frac{\Gamma\left(\frac{\rho}{\rho-1}\right)}{\pi}I_3(m^{\rho-1}/\Lambda^2) \right)    \,,
\end{multline}
\begin{multline}
    \chi_2^\text{dis}(m^{\rho-1}) \propto \\-m^{-2(\rho-1)}
    \left(       1    -  32A\,m^{5-4\rho}\frac{\Gamma\left(\frac{\rho}{\rho-1}\right)}{\pi}I_2(m^{\rho-1}/\Lambda^2)   \right)    \,.
\end{multline}
Note that when changing the variable from $\tau$ to $m$, the divergent integral $I_1$ cancels out.
The results reported in Sec. \ref{subsec:Critical exponents} can be obtained by defining
\begin{equation}
    u\equiv g\,m^{5-4\rho}\equiv 32\,A\,m^{5-4\rho}\,,
\end{equation}
where we remark that $A\propto1/M$ and thus $g$ is small in the limit $M\to\infty$. On the other hand, the dimensionless expansion parameter $u$ diverges as $\rho>5/4$ at the critical point $m\to0$ and this leads to the procedure developed in Sec. \ref{subsec:Critical exponents}.

\section{Conclusions}\label{sec:Conclusions}

In this work, we applied the $M$-layer construction to estimate the critical exponents of the spin glass in a field at zero temperature defined on a one-dimensional LR lattice. This application significantly extends the range of the method beyond its previous use on finite-dimensional hypercubic lattices in $D$ dimensions  \cite{angelini2020loop,angelini2022unexpected,angelini2025bethe,Angelini_2024Ising,angelini2025criticalexponentsspinglass}, proving its effectiveness also on more unconventional topologies. Notably, the key insight is that, once the loop expansion has been computed for a given model (in our case, a spin glass in a field at zero temperature), changing the underlying lattice requires only knowledge of the number of NBP between two lattice points. This observation significantly simplifies the analysis, making the $M$-layer approach an extremely useful tool for studying the same model on different topologies with minimal additional effort. While some models exist where the critical behavior is directly affected by the underlying lattice, in these cases, the application of this method is simple and could give quite interesting results.

Also, the consistency of the results strongly supports the method's validity. In particular, the result for the anomalous field dimension, $\eta=0$, is a non-trivial outcome that strongly depends on the topological loop expansion. Its emergence in our results supports the formalism's consistency and robustness. Remarkably, the scaling behavior observed aligns with the proposed correspondence between SR and LR exponents, which, from our computation, coincide at first order in the loop expansion, with the only exception of the correction-to-scaling exponents that differ below the upper critical dimension. This provides further evidence that the predictions of the $M$-layer construction are not only theoretically consistent but also physically meaningful.

The choice of the $M$-layer random graph and of the original lattice, {\it i.e.} the random graph with fixed connectivity and power-law weights,  was motivated by the possibility of making the analytical computations presented in this work. However, it is important to emphasize that the critical exponents belong to a {\it universality class} that we expect to include many different models, in particular, models with any finite  value of $M$, including notably $M=1$, and also random graphs with Poissonian distribution of the degrees, for which the probability of connection decreases with the distance as a power-law.
Our zero-temperature theory applies only to finite-connectivity models; and therefore, it cannot be directly applied to FC models with interaction strengths that decay as power-laws \cite{Kotliar_1983}.
As discussed in \cite{angelini2025criticalexponentsspinglass} a crucial point is whether the zero-temperature fixed point that we found is attractive on the finite temperature critical line. This would considerably enlarge the universality class, including models at finite temperature and thus also the model of Ref. \cite{Kotliar_1983}.

Finally, the most impactful contribution of this work concerns the critical exponents themselves. The model defined on this one-dimensional LR lattice is particularly well-suited for numerical simulations, due to its relatively simple structure. Meanwhile, our analysis yields the first (albeit approximate) analytical estimates of the critical exponents below the upper critical dimension for this system. These estimates can serve as valuable benchmarks for future finite-size scaling analyses from simulations. In fact, a natural next step would be to conduct a numerical investigation to test the predictions presented here. Confirmation would represent a significant progress in the field, offering a solid foundation for further analytical or numerical explorations of related systems.

\begin{acknowledgments}
We thank M. A. Moore and M. Weigel for useful discussions and R. A. Baños, L. A. Fernandez, V. Martin-Mayor and A. P. Young for sharing the code used in this paper. TR has been supported by funding from the 2021 first FIS (Fondo Italiano per la Scienza) funding scheme (FIS783 - SMaC - Statistical Mechanics and Complexity) from Italian MUR (Ministry of University and Research).
\end{acknowledgments}

\bibliographystyle{unsrt}
\bibliography{biblio}

% The \nocite command causes all entries in a bibliography to be printed out
% whether or not they are actually referenced in the text. This is appropriate
% for the sample file to show the different styles of references, but authors
% most likely will not want to use it.
%\nocite{*}

\end{document}